\documentclass[sigplan,screen]{acmart}

\AtBeginDocument{%
  \providecommand\BibTeX{{%
    \normalfont B\kern-0.5em{\scshape i\kern-0.25em b}\kern-0.8em\TeX}}}

\setcopyright{rightsretained}
\acmPrice{}
\acmDOI{10.1145/3486609.3487205}
\acmYear{2021}
\copyrightyear{2021}
\acmSubmissionID{splashws21gpcemain-p24-p}
\acmISBN{978-1-4503-9112-2/21/10}
\acmConference[GPCE '21]{Proceedings of the 20th ACM SIGPLAN International Conference on Generative Programming: Concepts and Experiences}{October 17--18, 2021}{Chicago, IL, USA}
\acmBooktitle{Proceedings of the 20th ACM SIGPLAN International Conference on Generative Programming: Concepts and Experiences (GPCE '21), October 17--18, 2021, Chicago, IL, USA}

\usepackage{booktabs}   
\usepackage{subcaption} 
\usepackage{tikz}
\usepackage{amsmath}
\usepackage{amsfonts}
\usepackage{graphicx}
\usepackage{listings}
\usepackage{array}
\usetikzlibrary{automata,positioning}
\usetikzlibrary{shapes.geometric, arrows}
\usepackage{multirow}
\usepackage[font=footnotesize,skip=2.5pt]{caption}
\usepackage{filecontents}
\usepackage{subcaption}
\usepackage{titlesec}
\usepackage{lipsum}
\usepackage{mathpartir}
\usepackage{enumitem}
\usepackage{balance}
\usepackage{microtype}
\usepackage{etoolbox}
\newtoggle{extend}
\toggletrue{extend} 
\newcommand{\figref}[1]{Fig.~\ref{#1}}  
\newcommand{\secref}[1]{Sec.~\ref{#1}}

\newcommand{\appref}[1]{Appendix~\ref{#1}}
\newcommand{\tbref}[1]{Table~\ref{#1}}

\newcommand{\CODE}[1]{\texttt{#1}}

\newcommand{\eg}{\textit{e.g.}}
\newcommand{\ie}{\textit{i.e.}}

\newcommand{\etc}{\text{etc}}

\newcommand{\emp}{\ensuremath{\cdot}}

\lstdefinelanguage{Scala}%
{keywords=[1]{abstract,%
  case,catch,char,class,%
  def,else,extends,final,finally,for,%
  if,import,implicit,%
  match,module,%
  new,null,%
  object,override,%
  package,private,protected,public,%
  for,public,return,super,%
  this,throw,trait,try,type,%
  val,var,%
  with,while,until,%
  yield,%
  assert,%
  Int, %
  println, %
  abstrct, %
  Range,%
  },%
  keywordstyle=[1]\color{violet},%
  keywords=[2]{ Rep,%
  },%
  keywordstyle=[2]\color{PigBlue},%
  keywords=[3]{@sec, %
  @bound,%
  },%
  keywordstyle=[3]\color{blue},%
  sensitive,%
  morecomment=[l]//,%
  morecomment=[s][\color{blue}]{@}{\ },%
  morecomment=[l][\color{blue}]@bound,%
  morecomment=[s]{/*}{*/},%
  morestring=[b]",%
  morestring=[b]',%
  showstringspaces=false%
}[keywords,comments,strings]%

\definecolor{dkgreen}{rgb}{0, 0.6, 0}
\definecolor{gray}{rgb}{0.5, 0.5, 0.5}
\definecolor{mauve}{rgb}{0.58, 0, 0.82}
\definecolor{violet}{rgb}{0.53, 0.0, 0.69}
\definecolor{violet2}{rgb}{0.93, 0.51, 0.93}
\definecolor{backcolour}{rgb}{0.95, 0.95, 0.92}
\definecolor{PigBlue}{RGB}{42, 0, 255}
\definecolor{depmap}{HTML}{00007B}

\newcommand{\keyw}[1]{\textsf{\textbf{#1}}}
\newcommand{\etp}[4]{#1; #2 \vdash #3 : #4}
\newcommand{\shi}{\ensuremath{\top}}
\newcommand{\slo}{\ensuremath{\bot}}
\newcommand{\sle}{\ensuremath{\sqsubseteq}}
\newcommand{\scup}{\ensuremath{\sqcup}}
\newcommand{\scap}{\ensuremath{\sqcap}}
\newcommand{\den}[1]{\llbracket #1 \rrbracket}
\newcommand{\stp}[5]{#1; #2\ [#3] \vdash #4 \triangleright #5}
\newcommand{\dtp}[2]{#1 \vdash #2}
\newcommand{\ptp}[1]{\vdash #1}

\newcommand{\nonterm}[1]{\textrm{\textit{#1}}}
\newcommand{\SYNTAXCATEGORY}[1]{\nonterm{#1}}
\newcommand{\Expr}{\SYNTAXCATEGORY{e}}
\newcommand{\Stmt}{\SYNTAXCATEGORY{s}}

\newcommand{\TYPEJUDG}[2] {{\Gamma} \vdash {#1}: {#2}}
\newcommand{\TYPEJUDGLIST}[3]{{#1} \vdash {#2} \triangleright {#3} }
\newcommand{\STMTTYPEJUDG}[2] {{\Gamma} \vdash {#1} \triangleright {#2}}
\newcommand{\fun}[1]{\textit{#1}}
\newcommand{\DOM}{\fun{dom}}
\newcommand{\SUBTYPE}{\ensuremath{\leq:}}
\begin{document}

\iftoggle{extend}
{
  \title[HACCLE: Metaprogramming for Secure Multi-Party Computation]{HACCLE: Metaprogramming for \\ Secure Multi-Party Computation - Extended Version}        
}{
  \title[HACCLE: Metaprogramming for Secure Multi-Party Computation]{HACCLE: Metaprogramming for \\ Secure Multi-Party Computation}   
}

\author{Yuyan Bao}
\email{yuyan.bao@uwaterloo.ca}
\authornote{Both authors contributed equally to this work.}
\authornote{Work performed while author was at Purdue University.}
\affiliation{
  \institution{University of Waterloo}
  \country{Canada}
}
\author{Kirshanthan Sundararajah}
\authornotemark[1]
\email{ksundar@purdue.edu}
\affiliation{
  \institution{Purdue University}
  \country{USA}
}
\author{Raghav Malik}
\email{malik22@purdue.edu}
\affiliation{
  \institution{Purdue University}
  \country{USA}
}
\author{Qianchuan Ye}
\author{Christopher Wagner}
\author{Nouraldin Jaber}
\email{{ye202,wagne279,njaber}@purdue.edu}
\affiliation{
  \institution{Purdue University}
  \country{USA}
}
\author{Fei Wang}
\author{Mohammad Hassan Ameri}
\author{Donghang Lu}
\email{{wang603,mameriek,lu562}@purdue.edu}
\affiliation{
  \institution{Purdue University}
  \country{USA}
}
\author{Alexander Seto}
\author{Benjamin Delaware}
\author{Roopsha Samanta}
\email{{aseto,bendy,roopsha}@purdue.edu}
\affiliation{
  \institution{Purdue University}
  \country{USA}
}
\author{Aniket Kate}
\author{Christina Garman}
\author{Jeremiah Blocki}
\email{{aniket,clg,jblocki}@purdue.edu}
\affiliation{
  \institution{Purdue University}
  \country{USA}
}
\author{Pierre-David Letourneau}
\author{Benoit Meister}
\author{Jonathan Springer}
\email{{letourneau,meister,springer}@reservoir.com}
\affiliation{
  \institution{Reservoir Labs}
  \country{USA}
}
\author{Tiark Rompf}
\author{Milind Kulkarni}
\email{{tiark,milind}@purdue.edu}
\affiliation{
  \institution{Purdue University}
  \country{USA}
}

\renewcommand{\shortauthors}{Bao and Sundararajah, et al.}

\begin{abstract}
Cryptographic techniques have the potential to enable distrusting parties to collaborate in fundamentally new ways, but their practical implementation poses numerous challenges.
An important class of such cryptographic techniques is known as Secure Multi-Party Computation (MPC). Developing Secure MPC applications in realistic scenarios requires extensive knowledge spanning multiple areas of cryptography and systems.
And while the steps to arrive at a solution for a particular application are often straightforward, it remains difficult to make the implementation efficient, and tedious to apply those same steps to a slightly different application from scratch.
Hence, it is an important problem to design platforms for implementing Secure MPC applications with minimum effort and using techniques accessible to non-experts in cryptography.

In this paper, we present the  HACCLE (High Assurance Compositional Cryptography: Languages and Environments) toolchain, specifically targeted to MPC applications.
HACCLE contains an embedded domain-specific language Harpoon, for software developers without cryptographic expertise to write MPC-based programs, and uses \emph{Lightweight Modular Staging} (LMS) for code generation.  

Harpoon programs are compiled into acyclic circuits represented in HACCLE's Intermediate Representation (HIR) that serves as an abstraction over different cryptographic protocols such as secret sharing, homomorphic encryption, or garbled circuits.
Implementations of different cryptographic protocols serve as different backends of our toolchain.
The extensible design of HIR allows cryptographic experts to plug in new primitives and protocols to realize computation. And the use of standard metaprogramming techniques lowers the development effort significantly.

We have implemented Harpoon and HACCLE, and used them to program interesting applications (\eg, secure auction) and key computation components of Secure MPC applications (\eg,  matrix-vector multiplication and merge sort). We show that the performance is improved by using our optimization strategies and heuristics.
\end{abstract}



\maketitle

\vspace{-2ex}
\section{Introduction}
\label{sec:intro}

Secure Multi-Party Computation (MPC) enables a group of distrusting parties to jointly perform computation without revealing any participant's private data that they do not wish to share with others.
It has broad practical applications, \eg, Yao's millionaires problem ~\cite{yao1982protocols}, secure auctions~\cite{auction06,DBLP:journals/concurrency/HinkelmannJMRS11}, voting, privacy-preserving network security monitoring~\cite{sepia}, privacy-preserving genomics~\cite{wang2015efficient,jagadeesh2017deriving}, private stable matching~\cite{Doerner16}, ad conversion~\cite{Kreuter17}, spam filtering on encrypted email~\cite{gupta2017pretzel}, and privacy-preserving machine learning~\cite{chet}.
Secure MPC applications are generally realized as circuits communicating information -- both private and public -- among parties.

Although MPC techniques and protocols have seen much success in the cryptography community, it is still challenging to build practical MPC applications.
Executing cryptographic protocols is notoriously slow, due to the encryption and communication overhead.
The largest benchmark reported in Fairplay~\cite{malkhi2004fairplay} -- a secure two-party computation system -- was finding the median of two sorted input arrays containing ten 16-bit numbers from each party. Running the benchmark required execution of 4383 gates, and took over 7 seconds on a local area network.
While improving computing capabilities and network bandwidth, implementation techniques can contribute to 3-4 orders of magnitude improvements~\cite{evans2017pragmatic}.
These techniques include optimizations that reduce the number of gates and the depth of a circuit and reduce the computational costs of executing a cryptographic protocol.
However, such optimizations do not exist in general-purpose compiler frameworks.

While several MPC frameworks have been proposed~\cite{SPDZ,oblivC,scale-mamba,lu2019honeybadgermpc,aby,securenn,jiff,mpyc,zhang2013picco,ezpc,rastogi2014wysteria,liu2015oblivm,CBMC-GC,emp-toolkit,oblivC}, they either provide low-level cryptographic primitives or high-level abstractions like traditional programming languages, but not both.
The low-level frameworks provide high degrees of customized protocol execution, but the users are generally expected to be experts in either one or both of cryptography and optimizing circuits.
These MPC frameworks provide little or no type safety to prevent semantic errors,
and it is difficult to write applications in a way that is portable across different protocols.
The high-level frameworks provide traditional programming abstractions that hide the data-oblivious nature of secure computation from programmers.
But these frameworks are tied to only one or a few protocols and their compilation procedures -- from high-level abstractions to low-level primitives -- are not easy to extend to perform application-specific optimizations~\cite{oblivC}.

\paragraph{Contributions} The main intellectual contribution of this paper is a toolchain for developing Secure MPC applications called HACCLE (High Assurance Compositional Cryptography: Languages and Environments). Our framework contains an embedded domain-specific language (eDSL) \emph{Harpoon} for designing MPC-based applications, and uses standard metaprogramming techniques to lower the development effort.
Allowing seamless construction of MPC-based applications by software developers without expertise in advanced cryptography is the main purpose of providing such a high-level programming language.
A Harpoon program is compiled to an acyclic combinational circuit, which is described in a HACCLE Intermediate Representation (HIR).
HIR exposes the essential data-oblivious nature of MPC, and allows cryptography experts to experiment with new primitives and protocols.
Our framework also provides a specialized backend for estimating the resource usage (\eg, compute time and memory space) prior to execution.

\medskip\noindent
This paper makes the following specific contributions:
\vspace{-0.5ex}
\begin{itemize}[leftmargin=*]
\item \textbf{HACCLE Toolchain}: A compilation framework to build and execute MPC applications written in Harpoon -- an embedded domain-specific language (eDSL) in Scala based on the LMS metaprogramming and compiler platform.
\item \textbf{HACCLE Intermediate Representation (HIR)}: An extensible circuit-like intermediate representation tailored to abstract cryptographic primitives used in MPC.
\item \textbf{Optimization Strategies}: Methods for optimizing the MPC application by specialization as it flows through each stage of our HACCLE toolchain.
\end{itemize}

The rest of the paper is organized as follows.
\secref{sec:bkg} provides background on cryptographic protocols involved in Secure MPC and motivates the need for developing MPC-based applications.
We describe the key impediments for developing practical MPC applications with the example of secure auctions.
\secref{sec:haccle} illustrates components of our compiler and HIR.
\secref{sec:workflow} describes the HACCLE toolchain and associated workflow.
\secref{sec:optimization} describes the optimizations implemented in our compiler toolchain.
\secref{sec:evaluation} discusses our toolchain on three case studies in detail.
\secref{sec:relatedwork} summarizes related work and \secref{sec:conclusion} concludes the paper.
The HACCLE implementation is available online at: \\ \url{https://github.com/YuyanBao/HACCLE}.

\section{Motivating Example and Background}
\label{sec:bkg}
As an example of Secure MPC, consider online auctions. 
Online auctions have great practical importance and different models are widely used, \eg, by eBay, Google AdWords, and Facebook.
In general, a secure online auction works as follows.
Buyers place their sealed bids on items, and for each item, the highest bidder is chosen to buy it.
In this setup, parties are not permitted to know others' bids. 
Hence, conducting successful secret auctions in the absence of a trusted authority requires cryptographic techniques to preserve the secrecy of bids while performing necessary computation, such as finding the highest bidder, in an assuredly trustworthy way.
One of the significant use cases of secure auctions is procurement via a competitive bidding process, where no participant trusts each other, including the auctioneer.
While a trusted third party handling the auction may be acceptable when the items under auction have low value, this is generally a less desirable option in high-value and corruption-prone environments, such as procurement for public construction contracts.

There are many different types of auction policies studied by economists and game theorists.
An auction where the highest bidder is chosen to buy the item by paying the highest bid is known as a {\em first-price} auction.
A {\em second-price} or Vickrey auction~\cite{vickrey61} is an alternative auction policy where the highest bidder is chosen to buy the item at the {\em second} highest price.
Second-price auctions provide buyers with the incentive to bid their true valuation and do not allow for price discovery (\ie, no ramping up of prices).
Hence, second-price auctions are especially suitable for high-value low-trust environments, such as public procurement.
Second-price auctions also apply to settings where multiple items are auctioned and/or bids may have additional structure, such as if/then conditions to evaluate specific contract terms that need to be taken into account for comparison.
Such settings are described as generalized second-price auctions.
Given that secrecy of the bids is preserved, the computation required when a single item is auctioned is simpler than when multiple items are auctioned.
Hence it is desirable both from programmability and efficiency viewpoints that the online auction application is written once for the general case and gets automatically and correctly specialized for the desired number of items, number of bidders, comparison logic, \etc. 

Most implementation techniques for Secure MPC applications (\eg, first- and second-price auctions) are based on circuits.
Equivalent functionality can be expressed as a Scala program: \eg, the following expresses an AND gate template, with bit-width determined by the input array:
\vspace{-1.2ex}
\begin{lstlisting}
val input = Array(0, 1, 1, 0)
var res = input(0)
for (i <- (1 until input.length))
  res = res & input(i)
res
\end{lstlisting} 
\vspace{-1ex}
Just like in DSLs for hardware design~\cite{chisl,spatial}, using metaprogramming techniques to \emph{stage} bitwise operations rather than execute them directly is the key to our approach.
Implementing \emph{secure} circuits then amounts to specializing the encoding and operators for the respective cryptographic backends.

We use Lightweight Modular Staging (LMS)~\cite{DBLP:conf/gpce/RompfO10} to turn the encoding and operators into \emph{staged} expressions, so that programs like the previous AND template become circuit generators. In LMS, type constructor \CODE{Rep[T]} is used to denote a \emph{staged} expression, which will cause an expression of type \CODE{T} to become part of the generated program.
The following code shows the high-level design of HACCLE intermediate representation (HIR) using LMS. 
The case classes \CODE{Bit} and \CODE{Num} define the primitive constructs of encoding boolean circuits and arithmetic circuits respectively, where the types \CODE{Rep[SBit]} and \CODE{Rep[SNum]} denote the staged representations of secure bits and numbers (see \secref{sec:ir}).
\vspace{-0.5ex}
\begin{lstlisting}
// Boolean circuit interface
abstract class SBit
case class Bit(val value: Rep[SBit], ... ) {
  def &(that: Bit) = { ... }
  def |(that: Bit) = { ... }
}
// Arithmetic circuit interface
abstract class SNum
case class Num(val value: Rep[SNum], ...) {
  def +(that: Num) = { ... }
  def -(that: Num) = { ... }
  def <(that: Num) = { ... }
}
\end{lstlisting}
\vspace{-0.5ex}
It is of course possible to implement \CODE{Num} on top of binary circuits and \CODE{Bit} arrays using standard half adders and full adders (see \secref{sec:ir}), but some secure cryptographic protocols directly support arithmetic circuits.

Now to express a secure first-price auction, we can use operations on an array of pairs of \CODE{Num}s that denote encrypted bidders' identities and their bids:
\vspace{-0.5ex} 
\begin{lstlisting}
// assume input: Array[(Num, Num)]
var res = input(0)
for (i <- (1 until input.length))
  res = if (res._2 < input(i)._2) input(i) else res
res
\end{lstlisting}
Observe that the linear sequence of operations in the above code results in a suboptimal circuit. Rewriting the code in a functional style, as, \CODE{input.reduce(\_  max \_)}, allows us to abstract over the reduction pattern and substitute the linear sequence with a tree reduction patten, which yields a circuit of logarithmic depth, allowing efficient parallel computation. Using known techniques for extracting functional dependencies from imperative loops~\cite{rompf2017functional,essertel2019precise}, this transformation is  automated and applied to for loops. Now, all we need are generic functions: \CODE{max}, \CODE{sndmax} (shown below) and \CODE{reduce} (\figref{fig:reduce}). The latter divides the computation into subproblems of size $n/2$ and call the subproblems recursively.
\begin{lstlisting}
// compare (bid id, bid value)
def max(a: (Num, Num), b: (Num, Num)): (Num, Num) =
  if (a._2 < b._2) b else a 
// compare (bid id, bid value, price = 2nd highest bid)
def sndmax(a: (Num, Num, Num), b: (Num, Num, Num)) =
  val prz = ... // 2nd highest of a._2,a._3,b._2,b._3
  if (a._2 < b._2) (b._1,b._2,prz) else (a._1,a._2,prz)
\end{lstlisting}
\vspace{-0.5ex}
Type classes, \eg, \CODE{Ordering[T]} or \CODE{Encoding[T]}, can be used to further abstract over comparison or access logic.

\begin{figure}[t]
\begin{lstlisting}
def reduce[T](input: Array[T])(f: (T, T) => T): T = {
  def rec(elems: Array[T]): T =
    if (elems.length == 1) elems(0)
    else {
      val b1 = elems.slice(0, elems.length/2)
      val b2 = elems.slice(elems.length/2, elems.length)
      f(rec(b1), rec(b2))
    }      
   rec(input)
}
\end{lstlisting}
\caption{Generic function \CODE{reduce} yielding a circuit of logarithmic depth.}
\label{fig:reduce}
\vspace{-1.5em}
\end{figure}
With the given comparator functions, we can transform the previous imperative code to a functional style, which generates optimal circuits:
\vspace{-0.5ex}
\begin{lstlisting}
// compute first-price auction
val max = reduce(input)(max)
// compute second-price auction
def initPrice(x) = (x._1, x._2, x._2)
def dropSecretBidValue(x) = (x._1, x._3)
val r = reduce(input.map(initPrice)))(sndmax)
val snd_max = dropSecretBidValue(r)
\end{lstlisting}
\vspace{-0.5ex}
For second-price auctions, we transform each element in the array to a 3-tuple of bidder's identity, highest bid, and initial price, and reduce with \CODE{sndmax}.
\secref{sec:auctions} shows a full implementation in our HACCLE toolchain.

We continue our discussion of Secure MPC background by looking at different protocols for secure computation, system models, trust models, and the offline/online paradigm.

\paragraph{Secret sharing.}
Secret sharing~\cite{secrectshare79} is a cryptographic technique that distributes secret data amongst a group of parties, and allows the secret to be reconstructed only when a sufficient portion of shares are combined.
A $(t, n)$-secret sharing scheme allows the secret $s$ to be split into $n$ shares.
Any $t-1$ of the shares reveal no information about $s$, while any $t$ shares can complete reconstruction of the secret $s$.

The SPDZ~\cite{SPDZ} and HoneyBadgerMPC~\cite{lu2019honeybadgermpc} frameworks serve as our secret sharing backends and provide Python-style programming environments for writing custom MPC programs.
These frameworks let developers express MPC programs (\eg, second-price auction) as arithmetic expressions.
Constructing the most efficient MPC programs is the major challenge for developers.
First, developers must know how to build an efficient circuit, \eg, realizing a balanced tree reduction to reduce the depth of a circuit and to parallelize the computation, instead of performing a linear reduction over a list of elements.
Second, developers must have a good understanding of the cost of every primitive operation (\eg, usage of logically similar but different comparison operators may yield different costs).
These challenges are significantly different from writing an efficient program in the traditional setting and can be successfully overcome by a compiler.

\paragraph{Homomorphic Encryption.}
Cloud computing may violate privacy.
In this scenario, one party wants to perform computation by outsourcing to another (possibly untrusted) party, \eg, training machine learning models of private data on a public cloud server.
This can be achieved by homomorphic encryption, another important cryptographic primitive.
{\em Homomorphic encryption} enables operations on encrypted data.  
The PALISADE~\cite{palisade}, TFHE~\cite{TFHE}, and HElib~\cite{Helib} libraries serve as our FHE backends.
They all implement asymmetric protocols that use a pair of public and private keys for encryption and decryption.
The TFHE library implements a very fast gate-by-gate bootstrapping mechanism~\cite{chillotti2016faster, chillotti2017faster}, and allows to evaluate a boolean circuit composed of binary gates over encrypted data.
The HElib library implements many optimizations to make homomorphic evaluation run faster.
The PALISADE library supports the BGV~\cite{bgv}, BFV~\cite{bfv1, bfv2}, and CKKS~\cite{ckks} schemes.
In cryptography, ciphertext and plaintext mean private and public information, respectively. In this paper, we may use these terms interchangeably.

\paragraph{Garbled Circuits.}
Yao's Garbled circuits \cite{yao1982protocols} is a two-party secure computation scheme for boolean circuits against semi-honest adversaries.
Obliv-C \cite{oblivC} is the library that we use to support Yao's Garbled Circuits protocols.

\paragraph{System and Communication Models.}
There are two popular system models for multi-party computation.
The MPC-as-a-service setting allows some parties to play the role of servers and to provide MPC services to clients with private input.
The other setting is where the parties running the MPC protocols are the participants who provide the input.
The HACCLE toolchain does not enforce a specific setting; instead, users choose the suitable setting for their applications and keep that setting in mind when developing programs.
Similarly, the HACCLE toolchain does not enforce any communication model.
The parties/machines could be fully connected, could form a star network structure, or could be any specified structure.
As long as the network structure is supported by one of HACCLE's backends, HACCLE is able to compile the program.

\paragraph{Trust/Adversary Models.}
Developing MPC applications requires understanding the security assumptions of an MPC library, such as the trust/adversary models.
There are two major adversary models: semi-honest and malicious. A {\em semi-honest} adversary follows the protocol, but tries to learn from received messages.
A {\em malicious} adversary has the same power as a semi-honest one in analyzing the protocol execution.
In addition, it may also control, manipulate, or arbitrarily inject messages to the network.
In HACCLE, programmers only need to provide a model of choice and the toolchain will pick proper sub-protocols to build up the MPC programs satisfying the adversary model described.

\paragraph{Offline Phase.}
The offline/online paradigm is applied by many MPC protocols and frameworks. The online phase uses a buffer of preprocessed input-independent values created during the offline phase. Thus, the MPC framework can run the offline phase to prepare them beforehand.
The online phase is where clients/users provide their inputs and get expected output; it can gain a significant speed-up with the help of the offline phase.
A number of preprocessed values are required for multiplications and comparisons.
The volume of preprocessing data depends on the online phase, and it is hard for programmers without security expertise to work out those requirements.
In HACCLE, programmers need not care about the secret parameters. 
They describe only the computation and the private information.
The HACCLE toolchain can synthesize suitable settings for the offline phase.

\section{Compiler}
\label{sec:haccle}
The HACCLE toolchain uses LMS~\cite{DBLP:conf/gpce/RompfO10} to support our towers of abstractions. Staging is a technique for building extensible, flexible DSLs by providing \emph{code generators} that successively lower higher-level abstractions to lower-level abstractions, and, ultimately, to executable code. Importantly, staging allows optimization to be performed at every level of the lowering process.
Hence, some optimizations can be performed at high levels of abstraction (\eg, optimization on plaintext computation (see \secref{sec:optimization})), while other optimizations can be performed at lower levels of abstraction. 
As a result, abstraction penalties are minimized. Another benefit of staging is that because the translation is written in terms of generators, it is simple to add new abstractions at any given level.

\subsection{Staged Compilation}\label{sec:staging}
\begin{figure}[t]
  \centering
\includegraphics[scale=.3]{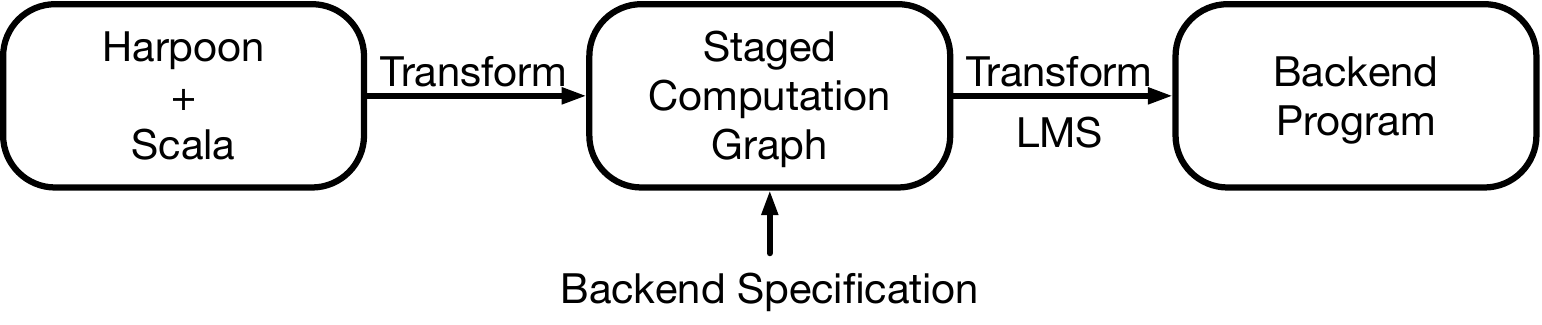}
\caption{Compilation in HACCLE.}
\label{fig:compilation}
\vspace{-1.5em}
\end{figure}

Multi-Stage Programming~\cite{DBLP:journals/tcs/TahaS00} (or staging) is the programming language technique that executes programs
in multiple stages.
A staged computation does not immediately compute a result, but returns a program fragment that represents the computation and that can be explicitly executed to form the next computational stage.
The key benefit of staging is that the present-stage code can be written in a high-level style,
yet generates future-stage code that is very low-level and efficient.
\figref{fig:compilation} illustrates an end-to-end compilation path in HACCLE. The compiler takes a Scala program with Harpoon annotations (see \secref{sec:harpoon}), and constructs a computation graph that expresses an abstract circuit. Given a backend specification, the compiler will generate a target program for it. Currently, our compiler is not able to automatically choose an appropriate backend and initialize all the parameters for it. Thus, a backend specification is needed. It is a file that contains a set of parameters for translating an abstract circuit to a concrete backend program. 

\paragraph{Generative Programming and Lightweight Modular Staging (LMS).}
As mentioned in \secref{sec:bkg}, the HACCLE compiler uses LMS for code generation due to its metaprogramming capabilities, and the type constructor \CODE{Rep[T]} is used to denote a staged expression. For example, the type \CODE{Rep[SNum]} denotes an encrypted integer. Given two \CODE{Rep[SNum]} values \CODE{a} and \CODE{b}, \emph{evaluating} the expression $a + b$ will \emph{generate} code for a given backend. For the Helib backend, the generated code will be \CODE{Ctxt r = a; r += b;}, where \CODE{Ctxt} is the type of a ciphertext in the Helib library. For the TFHE backend, the generated code will be:
\vspace{-1ex}
\begin{lstlisting}[language=C,basicstyle={\footnotesize\ttfamily}]
LweSample* x5 =
new_gate_bootstrapping_ciphertext_array(64, x2->params);
fhe_add(x5, a, b, 64, bk);
\end{lstlisting}
\vspace{-1ex}
where \CODE{LweSample} is the type of a ciphertext in the TFHE library. 
As a TFHE program does not provide arithmetic expressions and operations, 
the compiler encodes an integer as a bit-array of size 64. The function \CODE{fhe}\_\CODE{add} is part of our HACCLE library of the TFHE library.

\subsection{Harpoon}
\label{sec:harpoon}
HAccle Rich Representation for Program OperatiON ({\em Harpoon}) language is an expressive subset of Scala for writing MPC programs. It is an imperative and monomorphic language, featuring standard control flow operations: loops,
function calls, conditionals, and recursions. The language is designed to be expressive enough that programmers could easily write Harpoon code directly, while being constrained enough to ensure that Harpoon programs can be implemented via translation to secure low-level computation. In practice, Harpoon serves as the top-level IR for the HACCLE pipeline, and is the language for end-user programs.

The Harpoon language is not only able to access Scala libraries, but also provides a set of cryptographic data structures, \eg,  \CODE{HArray[T]} is an encrypted array that allows one to index on ciphertexts. 
It also provides a set of security annotations that are read via reflection and are used to direct code generation.
They are agnostic to the target backend, and are used by subsequent stages of the HACCLE pipeline.
For example, the annotation \CODE{sec} is used to mark the provider (also the owner) of private data.
Recursive functions and loops may be annotated with an upper bound on the number of recursive calls and iterations. 
This expression can reference the parameters of the function, allowing this bound to vary according to the context where a function is called, \eg, consider the signature of \CODE{merge} function:
\vspace{-0.5ex}
\begin{lstlisting}
@bound(a.length + b.length)
def merge(a: HArray[Int], b: HArray[Int]): Harray[Int]
\end{lstlisting}
\vspace{-0.5ex}
The upper bound of the number of recursive calls is the sum of the length of the two input arrays. Note that the semantics of function calls in Harpoon is not impacted by the bound; rather it is used by subsequent stages of the pipeline to bound the invocation of a recursive function call (see \secref{sec:obliv}).

The annotated program is also equipped with a type system, and ensures that information about private data cannot be leaked. This provides the first-layer guarantees that the programs can be successfully compiled by the later stages of the pipeline. Consider the statement \CODE{println(a)}, where $a$ is annotated as private data. The compiler will report a type error, as encrypted data is not understandable or meaningful to users. But the assignment \texttt{@sec(alice) val r = a}
is permitted, as the annotation expresses that the variable \CODE{r} stores encrypted data.
While the type system at this stage does not make use of fine-grained ownership information, this information will be passed down through the pipeline. \iftoggle{extend}{
See \appref{sec:syntax-harpoon} and \appref{sec:HarpoonTyping} for the details of Harpoon syntax and typing rules.}{See~\cite{haccle} for the details of Harpoon language.}

\subsection{Intermediate Representation}
\label{sec:ir}
HACCLE intermediate representation (HIR) serves as an interface between high-level programming languages and cryptographic backends.
HIR is a domain-specific intermediate language, and gains benefits from LMS to support towers of abstractions. It encompasses all the primitive operations which we have supported so far, \eg, encryption, decryption, sharing, and combining.
\paragraph{Multi-level IR.}
Different backends may support different sets of operations in HIR---no backend is ``complete'' in that there is a direct implementation of each HIR operation in that backend. For example, the TFHE backend provides a complete set of logical operations, but does not support arithmetic operations. In contrast, other backends may support arithmetic operations but not boolean operations. The compiler's job is to {\em rewrite} HIR circuits to be compatible with backends.

\begin{figure}[t]
  \centering
  \small
  \begin{minipage}{0.45\textwidth}
    \centering
    \begin{tikzpicture}[squarednode/.style={rectangle, draw=blue!60, very thick, minimum size=3mm},squarednode1/.style={rectangle, draw=blue!60, very thick, minimum size=3mm, text centered, text width=2.5cm},]
     \node[squarednode] at (-4, 0) {\small Float, FloatArray};
     \node[squarednode] at (-4, -1) {\small UNum, UNumArray};
     \node[squarednode] at (-1, -1) {\small Num, NumArray};
     \node[squarednode] at (-2, -2) {\small Bit, BitArray};
     \draw[thick, ->](-4, -0.3) -- (-4, -0.7);
     \draw[thick, ->](-4, -1.3) -- (-2, -1.7);
     \draw[thick, ->](-1, -1.3) -- (-2, -1.7);
    \end{tikzpicture}
  \end{minipage}
  \caption{Example of multi-level HIRs.}
  \label{fig:hir-nodes}
  \vspace{-2em}
\end{figure}
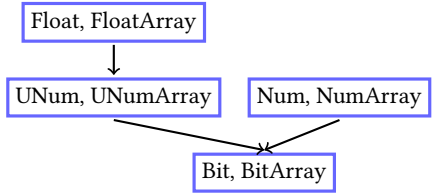

As shown in \figref{fig:hir-nodes}, HIR is a multi-level IR. The compiler can thus use rewrites to target the subset of operations that a given backend supports. For example, arithmetic operations (adds, multiplies) can be rewritten into bit-level implementations (as, \eg, ripple-carry adders, or bit-level implementations), or boolean operations can be represented as arithmetic operations that happen to operate over $\mathbb{Z}_2$. We have developed a set of these rewrite rules for various backends (and, indeed, rely on exactly this type of rewrite to support floating point operations).

A key task for integrating a new backend is identifying what set of HIR operations that module supports, hence directing the compiler to perform appropriate rewrites. Notably, if the compiler {\em cannot} rewrite an HIR circuit to target the set of operations a backend supports, it will manifest as a type error, providing feedback to the user.

In the scenario of using a FHE scheme, an integer is encoded as the \CODE{Num} data structure shown below, where the fields \CODE{provider} and \CODE{value} are abstraction of the party who provides the value and the encrypted value respectively.
\vspace{2em}
\begin{lstlisting}
case class Num(
  val provider: Set[Rep[SOwner]], // who provides it
  val value   : Rep[SNum]         // encrypted value
)
\end{lstlisting}
\vspace{-0.5em}
In this case, a variable declaration statement in Harpoon, \ie, 
\texttt{@sec(alic) val x = 5;},
is transformed to \CODE{val o = new Owner(alice);  val x = Num(o, 5);} in HIR.

In the scenario of using a secret sharing scheme, an integer is encoded as the \CODE{ShareNum} data structure in HIR shown below. It expresses a general secret sharing protocol. The \CODE{provider} is the one who contributes the \CODE{value} that is shared among a set of \CODE{players} with \CODE{threshold}. The set of \CODE{observers} are those who are allowed to access the value once it gets combined.
\vspace{-0.5ex}
\begin{lstlisting}
case class ShareNum(
  val provider  : Set[Rep[SOwner]], // who provides it
  val players   : Set[Rep[SOwner]], // players
  val observers : Set[Rep[SOwner]], // who observes it
  val threshold : Int,              // threshold
  val value     : Rep[SShareNum]    // shares
)
\end{lstlisting}
\vspace{-0.5ex}
In addition, HIR provides libraries for implementing secure computation. Those libraries are not supported by general-purpose compilers, but are essential to build interesting multi-party applications with security guarantees.
For example, the following shows the operations of an array supporting indexing on a ciphertext, where \texttt{arr} is an HIR array. 
\begin{itemize}[leftmargin=*]
  \item \lstinline|arr(i)|: array index, where $i$ is a plaintext or a ciphertext.
  \item \lstinline|arr.update(i, v)|: update the $i$th element with the value $v$, where $i$ is either a plaintext or a ciphertext.
  \item \lstinline|arr.slice(i, j)|: array slicing from the $i$th element until the $j$the element, where $i$ and $j$ are plaintext.
  \item \lstinline|arr.length|: the length of the array
\end{itemize}

The way these array operations with secure indices are currently implemented is through, essentially, a naive Oblivious RAM (ORAM): to index into an array with a ciphertext index, the compiler generates a circuit that wires every array element, and a secure selector (multiplexer) to output the desired array element. This is equivalent to a set of \emph{if-then-else}s to choose the desired array element, except with a logarithmic depth instead of a linear depth. Writing to an array element with a ciphertext index is the equivalent of an array copy, where each element of the new array performs a check for whether the old element of the array should be copied, or the ``update'' value should be copied.

As implementation details of cryptographic backends are abstracted away from HIR, our framework can be easily extended to support more advanced cryptographic backends, for example, a backend with ORAM. Here, we would leverage HIR's ability to provide backend-specific rewrite rules, and would directly rewrite array operations to ORAM operations.

\paragraph{Type System.}
HIR also abstracts away the implementation details of cryptographic primitives and protocols. For example, an addition operation does not specify how a secure addition is achieved, as different protocols perform in different ways. But the type rules provide an approximation of data access policy that specifies how data is provided, accessed, and shared. For example, an addition operation on two shared numbers is only allowed on the same set of players with the same threshold, which are known at compile time. And the result is provided by either one of its operand's providers with the same set of players with the same threshold, and is allowed to be accessed by either one of the operands' observers.
\vspace{-0.5ex}
\begin{lstlisting}[basicstyle={\footnotesize\ttfamily}]
def +(x: ShareNum, y: ShareNum) = {
 assert(x.players.equals(y.players))
 assert(x.threshold == y.threshold)
 ShareNum(x.provider | y.provider, players,
 x.observers | y.observers, threshold, value.+(y.value))
}
\end{lstlisting}
\vspace{-0.5ex}
Given a cryptographic backend, HIR code is further transformed to a program with the corresponding cryptographic semantics. And the HIR type system is refined to provide more precise information on data access policy. For example, the type rule of the addition operation is refined to the following when using the additive secret sharing scheme.
\vspace{-0.5ex}
\begin{lstlisting}[basicstyle={\footnotesize\ttfamily}]
def +(x: ShareNum, y: ShareNum) = {
 assert(x.players.equals(y.players))
 assert(x.players.size == x.threshold)
 assert(x.threshold == y.threshold)
 ShareNum(x.provider | y.provider, players,
 x.observers & y.observers, threshold, value.+(y.value))
}
\end{lstlisting}
\vspace{-0.5ex}
The type rule checks it is a $n$-out-of-$n$ secret sharing scheme, \ie, \CODE{x.players.size == x.threshold}.
The refined type rule provides a stronger security guarantee, \ie, the transformed program is compatible with the semantics of the backend. For example, an FHE target program is not transformed to a program that may invoke secret sharing primitives. 
\iftoggle{extend}{
See \appref{sec:syntax-hir}, \appref{sec:hir-typing-rules} and \appref{sec:types-backends} for the details of HIR syntax and typing rules.}{See~\cite{haccle} for the details of HIR.}

\subsection{Obliviousness}
\label{sec:obliv}
In addition to bridging the semantic gap between a high and a low-level language, our compiler also bridges the semantic gap of obliviousness. 
A program without privacy concern diverts its control flow  according to the input: statements are executed conditionally, loop for a variable number of iterations, etc.
To protect privacy, boolean and arithmetic circuits have to be oblivious in the sense that they perform the same sequence of operations regardless of the input. 
The following transformations may seem quite inefficient at first sight, but they are absolutely necessary in order to maintain obliviousness.
\paragraph{Encrypted Array Indexing.}
Indexing an array with a ciphertext is encoded as a multiplexer circuit that takes every element of the array as an input and outputs the element in the position. 
This multiplexer circuit consists of integer comparators and selectors.

\paragraph{Conditional Execution.}
After a typed Harpoon program is transformed to HIR code, there are two types of if-constructs allowed. One is the standard if-statement, where its condition depends on plaintext comparisons, and the two branches consist of a sequence of statements that may have side effects.
The other has the form \texttt{z = if (b) x else y}, where the value of \texttt{b} is the result of private comparisons. Obliviousness is effectively guaranteed by executing both the consequent and alternative branches. If the backend is a boolean circuit, this if-construct is further transformed to a selector. If the backend is an arithmetic circuit, the program is transformed to \texttt{z = b * x + (1 - b) * y}. In the following Harpoon code snippet, the variable \texttt{arr} stores a sequence of shared numbers, and the comparison result of \texttt{max < arr(i)} is a shared secret value. Thus, the program 
\vspace{-0.5ex}
\begin{lstlisting}
if (max < arr(i)) { max = arr(i) }
\end{lstlisting}
\vspace{-0.5ex}
is transformed to
\vspace{-0.5ex}
\begin{lstlisting}
val b = max < arr(i)
max = b * arr(i) + (1 - b) * max
\end{lstlisting}
\vspace{-0.5ex}
Note that such a program transformation is non-trivial for a program allowing mutable states.
Currently, an if-statement will be transformed if the side effects of its two branches can be syntactically detected.

\paragraph{Loops and Recursion.}
All function calls are treated as macros and are simply inlined. 
All loops are unfolded as the number of iterations is a compile-time constant.
\figref{fig:gcd} demonstrates our treatment of recursive calls, where the obliviousness is achieved by using the extra plaintext parameter \CODE{d} on the right side of the figure. In the transformed program, the value \CODE{d} is initialized by the Harpoon annotation and decreases with each iteration. This makes sure that the recursive call only iterates \CODE{d} times. Note that the function \CODE{func} is a polymorphic overloading function in HIR.

\begin{figure*}[t]
\small
\begin{minipage}[t]{0.22\textwidth}
  Scala Program
\begin{lstlisting}[basicstyle={\footnotesize\ttfamily}]
val a = 5
val b = 15
def gcd(x: Int, y: Int)
  : Int = {
  if (x == 0) y
  else gcd(y % x, x)
}
println(gcd(a, b))
\end{lstlisting}
\end{minipage} %
\begin{minipage}[t]{0.3\textwidth}
  Harpoon Program
\begin{lstlisting}[basicstyle={\footnotesize\ttfamily}]
@sec(alice) val a = 5
@sec(alice) val b = 15
@bound(5)
def gcd(x: Int @sec, y: Int $\color{blue}{\texttt{@sec}}$)
  : Int @sec = {
  if (x == 0) y
  else gcd(y % x, x)
}
@reveal(alice) val r = gcd(a, b)
println(r)
\end{lstlisting}
\end{minipage} %
\begin{minipage}[t]{0.42\textwidth}
 HIR Program
\begin{lstlisting}[basicstyle={\footnotesize\ttfamily}]
val o = Owner(alice)
val a = Num(o, 5)    val b = Num(o, 15)
val gcd = func((d: Rep[Int], x: Rep[SNum], 
                y: Rep[SNum]) => {
  if (d == 0) y
  else if (x == 0) gcd(d-1, x, y)
       else gcd(d - 1, y % x, x)
})
val r = Num(o, gcd(5, a.value, b.value)).eval(o)
println(r)
\end{lstlisting}
\end{minipage}
  \caption{Compute the Greatest Common Divisor (GCD) of two numbers. The left one shows the Scala textbook implementation. The middle one shows the Harpoon program. The annotations express that a user, \CODE{alice}, computes the GCD of her private data \CODE{a} and \CODE{b} through a different party, which performs computation on the data in an encrypted form, and provides the encrypted results to \CODE{alice}. The right one shows the corresponding HIR program. The translated \CODE{gcd} function has one extra parameter \CODE{d} initialized by the bound Harpoon annotation, and decreases with each iteration.}
  \label{fig:gcd}
  \vspace{-2ex}
\end{figure*}

\subsection{Code Generation}
\paragraph{Cryptographic Backends.}
In the context of building circuits, LMS is used to specialized a circuit with respect to a target backend. The outcome of such a programmatic specialization is a compiled target of the circuit.
The code generator transforms an abstract circuit to a concrete one for a given backend. 
\iftoggle{extend}
{
  For example, the following adder expressed in HIR is specialized to the boolean circuit shown in \figref{fig:add-tfhe} and arithmetic circuit shown in \figref{fig:add-bgv} in the \appref{sec:appA}.
}{
  For example, the following adder expressed in HIR is specialized to a boolean or arithmetic circuit based on the backend.
}
\vspace{-0.5ex}
\begin{lstlisting}
val o1 = Owner();
output((Num(o1, 10).+(Num(o1, 5))).eval(o1))
\end{lstlisting}
\vspace{-0.5ex}
The essence of multi-stage programming is to generate efficient programs using high-level constructs without run-time penalty \cite{taha2004gentle}.
The example in \figref{fig:sum} a shows a code snippet that generates a for loop. Note that the \CODE{if} condition is composed of a plaintext boolean type, so this code is executed at code generation time as shown \figref{fig:sum} b.
\begin{figure}[!htp]
\vspace{-0.5ex}
\small
\begin{tabular}{l}
(a) HIR code example:\\
\begin{lstlisting}[basicstyle=\scriptsize\ttfamily]
val sum = func((x: Rep[SNumArray], len: Rep[Int]) => {
  var n = 0     val b = true
  var res = Num(o1, 0).value
  while (n < len) {
    if (b) {  res = res + x(n) }
    n += 1
  }
  res
})
\end{lstlisting}
\\
(b) Generated C code of TFHE backend:\\
\begin{lstlisting}[language=C,basicstyle=\scriptsize\ttfamily]
const LweSample* x3(const LweSample* x4, int x5){
 int x6 = 0;
 const LweSample* x7 = num_init(0, 64 ,x2);
 while (x6 < x5) {
  x7 = add(x7, array_index(x4, x6, 64, x0), 64, x0);
  x6 = x6 + 1;
 }
 return x7;
}
\end{lstlisting}
\end{tabular}
\caption{(a) HIR code example (b) Generated C code of TFHE backend.}
\vspace{-1em}
\label{fig:sum}
\end{figure}

\paragraph{Resource Estimation.}
\label{resource-estimation}
This is one of the special noteworthy backends: instead of performing a computation, it 
generates a graphical representation of the HIR circuit, which is fed to a generic ``Evaluator''. 
This is a {\em resource estimation program} that traverses the graph and performs analysis at each node.
The estimator is parameterized on a given resource model, which specifies costs of each node, edge type, and the depth of each edge in the graph.

At the most basic level, the resource estimation framework expects an enumeration of the \emph{abstract gates} for a particular cost model, a description of how each HIR node type affects these gates, and depths. 
The total cost is tallied in terms of abstract gates.
For example, a cost model for a secret sharing backend may have \emph{round complexity} and \emph{communication complexity} as its abstract gates, whereas a circuit backend may have AND, OR, and NOT as its abstract gates.
The evaluator traverses the HIR graph and accumulates the abstract gate costs produced by each node, and tracks the maximum total depth encountered for critical path estimation. 
In the case of a secret sharing scheme, traversing the graph will potentially increment round and communication complexity as new computation nodes are encountered, whereas a circuit backend will increment gate costs. 
These gate costs are then instantiated with specific costs (in terms of lower-level operations) based on the resource estimates determined by cryptographic experts.

This framework can also be easily extended to evaluate costs that do not follow this simple model. 
A data structure at each HIR node and a function that performs accumulation of cost based on the type of the node are sufficient to estimate the cost.
Cost models can also be parameterized on values which are configurable but known at compile time (\eg, integer bitwidth). The prime modulus can be determined by the security specifications, and specific edge costs.
The ability to estimate the cost of a program becomes useful when selecting a target from multiple backends.
A program may be better suited for execution on a particular backend than another.
If the available backends' cost models are comparable, then we can generate resource estimations to choose the best one for execution. 

\section{HACCLE Workflow}
\label{sec:workflow}
\begin{figure}[t]
    \centering
    \includegraphics[scale=.24]{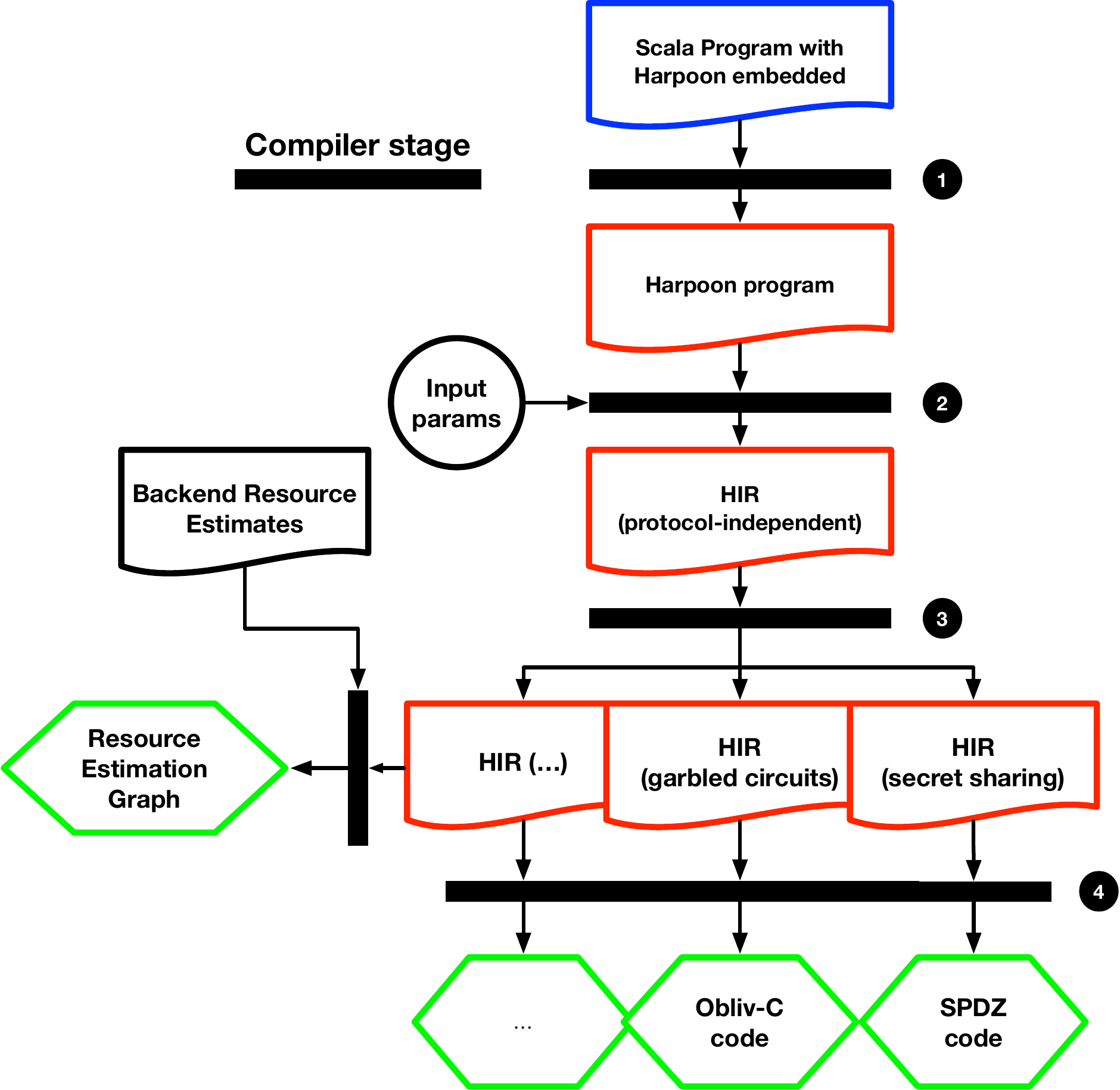}
    \caption{HACCLE Compilation Framework.}
    \label{fig:framework}
\end{figure}
This section describes the compilation flow of our HACCLE framework as shown in \figref{fig:framework}.
In the very first stage of the flow, an input program is staged to a complete Harpoon program that consists of an entry point for the inputs provided by the parties, computation and necessary revealing of results.
The Harpoon program is compiled to HIR code, which is one big acyclic circuit,
and is further lowered to the protocol specific HIR program.
Finally, the code for a specific backend is generated from the low-level HIR program.
Resource estimation models the resource usage of the computation in a specific protocol, and guides the compiler to generate optimal code.
The following subsections illustrate the stages of our toolchain from writing an MPC application as a program to executing it using different protocols, and how the type system provides various security guarantees at different stages.

\subsection{Specifying the Program}
A programmer starts by providing a Scala program that embeds a secure computation, which is written in Harpoon (see \secref{sec:harpoon}).
The Scala program runs at client locations, and is responsible for processing input, setting up communication channels, \etc.
The Harpoon program actually performs the secure computation that is written parametrically: effectively, a Harpoon program is a function that accepts the number of parties and their inputs as parameters.

\subsection{Generating a Circuit}
\paragraph{Stage 1}
The first stage of compilation transforms a Scala + Harpoon program to a pure Harpoon program, \ie, executing a Scala program stages away the {\em non-Harpoon} fragment of the code: local input files are read into memory and connections are set up to the relevant servers.

After the stage 1 compilation, a Harpoon program represents {\em just the secure computation that must be performed}. This program will eventually be transformed to a circuit that performs the desired secure processing. However, the secure computation is not ready for execution yet. Any publicly known information about the inputs (\eg, the bitwidths, or the maximum input size) has not yet been incorporated into the circuit, and the input values are not yet known.
At this stage, the Harpoon type system provides the key security guarantee that private data will not leak via public channels. 

An important note is that each Harpoon program represents a single secure computation that  compiles to a single circuit.
Hence, the Harpoon program must compile down to a circuit whose size is determined only by the publicly available information about the inputs.
In many applications, there are multiple secure computation that must occur (\eg, in database applications, there may be multiple queries; each query represents a different secure computation).
Here, we leverage the blurred distinction between compile time and runtime.
Generating a Harpoon program happens at what programmers traditionally consider {\em run time}: the Scala program is {\em actually running} to produce the Harpoon program.
Hence, the {\em Scala program} can include a loop over the set of queries, and for each query, a new Harpoon program is generated, compiled and executed.
The abstraction in Scala has no runtime overhead for the generated code
since it is executed at the Scala runtime,
offering the so-called ``abstraction without regret'' (see \secref{sec:staging}).

\paragraph{Stage 2}
The next step is to generate an {\em abstract circuit}: a Harpoon program is compiled down to HIR code (see \secref{sec:ir}), which is, essentially, a bounded-size and single-assignment representation of the program. Here, the bound annotation in the Harpoon program is used to unroll loops and inline recursive functions, leading to a functional and loop-free  representation of the program.
The HIR program at this stage is still independent of a particular protocol. 
Hence, it is essentially a direct translation of the Harpoon program into HIR code without considering the abilities of any particular backend.
The key typing guarantee that HIR code provides at this level is that the appropriate HIR operation will be used based on whether inputs to an operation are private or public. 

\paragraph{Stage 3}
The next compilation stage specializes an HIR circuit to a specific protocol. The choice of protocol is determined by the security specification file. Here, we do not change the {\em language} representation of the program---the resulting program is still in HIR. Instead, this stage rewrites HIR code to limit the use of HIR operations to those supported by a particular backend. For example, a backend that only supports boolean operations requires translating all operations on integers and floating point to bit-level operations. Similarly, a backend that only supports operations on integers requires translating floating point operations to decomposed operations on the component parts (mantissa and exponent).
Here, HIR switches to the use of backend-specific type systems that enforce the following property: a type-checked backend-specific HIR circuit enforces the requirements of that backend for security (\eg, the set of sharers matches up when performing operations in a secret-sharing backend).
\paragraph{Stage 4}
The final step of generating a circuit is specific to a backend implementation. Here, an HIR circuit is translated to be compatible with a particular backend. This is the key module interface provided by our system. It may require translating the circuit to a set of API calls (\eg, our TFHE backend), or to a different programming language (\eg, translating to Obliv-C for the garbled-circuit backend, or Scale-Mamba for the secret-sharing backend). The backend is configured based on the information in the security specification file.
At this point, the circuit is in an executable form, and can perform the desired secure computation, using the actual inputs from the various parties.

\section{Optimization}
\label{sec:optimization}
Our compiler contains a set of optimizing transformations, \eg, peephole optimizations, common subexpression elimination, constant folding, and dead code elimination. 
In addition to those optimizations that a general purpose compiler has, we identified several optimizations specific to Secure MPC circuits. 
Given an in-memory representation of a boolean or an arithmetic circuit, these optimizations reduce the depth of circuits and the number of costly gates.

\subsection{Scalar Multiplication}
\label{sec:privat-mutl}
The multiplicative depth of circuits is the main practical limitation in performing computation over encrypted data.
We identify that multiplication can be eliminated when one of the operands of a multiplication is 0 or 1 in plaintext.
In addition, consider the case of calculating $\mathit{pow}(x, n)$, where $x$ is an encrypted number.
The compiler can divide the computation into subproblems of size $n/2$ and call the subproblems recursively. 
\figref{fig:pow} shows the program of computing $\mathit{pow}(2, 8)$, where 2 is private. The Harpoon program is transformed to HIR code, and is further generated to the TFHE program, where the function \CODE{unum\_mul} multiplies two 64-bit encrypted numbers. The generated program only needs $O(\log n)$ multiplies.
This optimization is simple, but has a dramatic impact on performance.
\begin{figure}[t]
    \small
    \begin{minipage}{0.25\textwidth}
    \small Harpoon Program:
    \begin{lstlisting}[basicstyle={\ttfamily\footnotesize}]
    @sec(alice) val a = 2
    scala.math.pow(2, 8)
    \end{lstlisting}
    \end{minipage}%
    \begin{minipage}{0.25\textwidth}
    \footnotesize HIR Program:
    \begin{lstlisting}[basicstyle={\ttfamily\footnotesize}]
    val o1 = Owner()
    UNum(o1, 2).pow(8)
    \end{lstlisting}
    \end{minipage}
    \begin{minipage}{0.5\textwidth}
    \footnotesize Generated TFHE program:
    \begin{lstlisting}[language=C,basicstyle=\small\ttfamily]
    const LweSample* x3 = unum_init(2, 64, x2);
    LweSample* x4 = unum_mul(x3, x3, 64, x0);
    LweSample* x5 = unum_mul(x4, x4, 64, x0);
    return unum_mul(x5, x5, 64, x0);
    \end{lstlisting}
    \end{minipage}
    \caption{Computing $\mathit{pow}(2, 8)$, where $2$ is private, and \CODE{x0} and \CODE{x2} are the cloud key and private key used for encryption.}
    \label{fig:pow}
    \vspace{-2em}
\end{figure}

\begin{figure}[t]
    \centering
    \begin{subfigure}[b]{0.3\textwidth}
       \includegraphics[scale=.35]{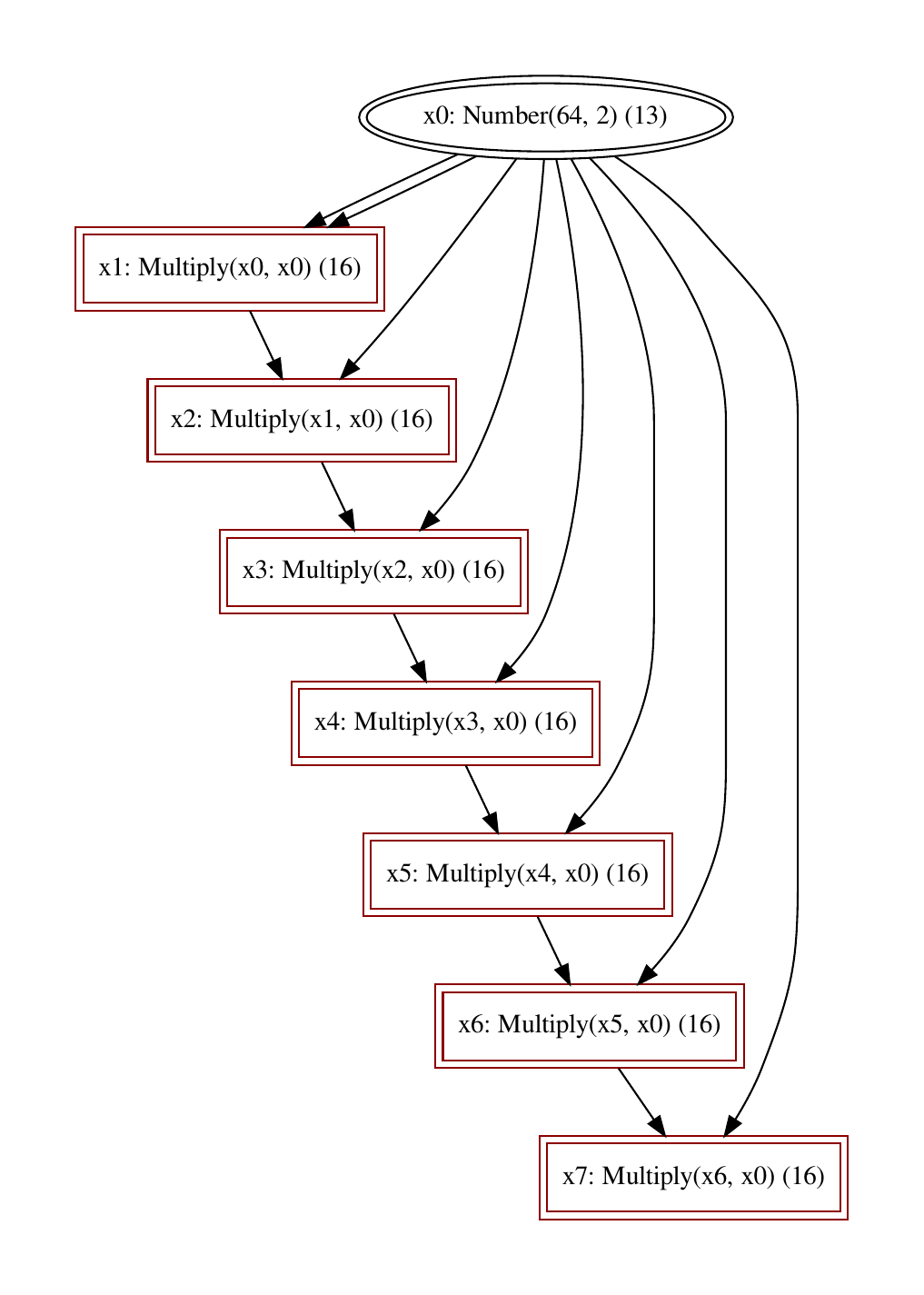} 
    \end{subfigure}%
    \begin{subfigure}[b]{0.3\textwidth}
       \includegraphics[scale=.35]{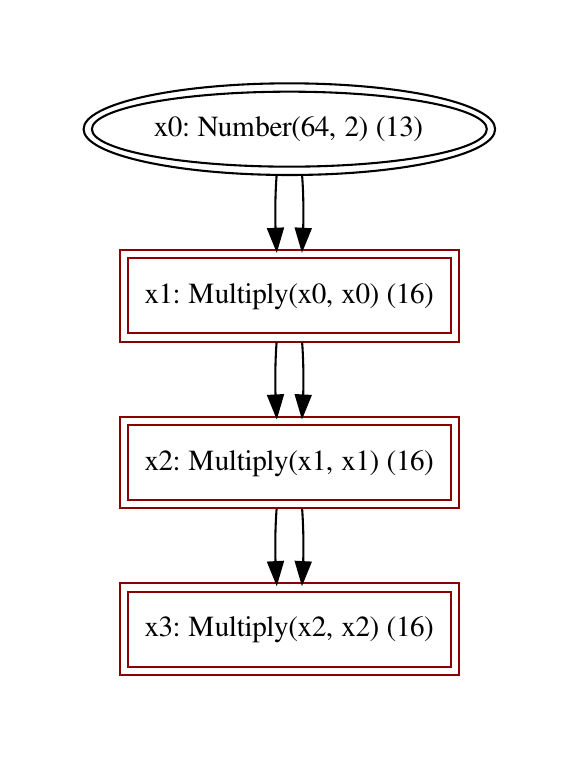} 
    \end{subfigure}
    \caption{Graphs of computing $\mathit{pow}(2, 8)$: before (left) and after applying optimizations (right). }
    \label{fig:pow-graph}
    \vspace{-1em}
\end{figure}

The effectiveness of the optimization is clearly demonstrated in \figref{fig:pow-graph}, which shows the graphs of the generated circuits. The left (before optimization) is a depth-7 circuit with 7 multiply gates. The right (after optimization) is a depth-3 circuit with three multiply gates.

The generated graphs show an abstract model of execution cost where each operation is treated as atomic. However, the resource estimation framework can be specialized to particular backends by providing the corresponding models of execution cost (in terms of communication complexity, number of logic gates, etc.).
These backend-specific resource estimates can be used to compare different optimization strategies and intelligently select the appropriate one based on the execution semantics of the targeted backend.
In addition, as mentioned in \secref{resource-estimation}, these specialized estimates even let us pick the most optimal backend to target.

\subsection{Private Comparison}
\label{sec:private-comp}

The private comparison is a major bottleneck in MPC protocols due to their inherent non-arithmetic structure \cite{couteau2016efficient}.
Private comparison operators include $<$, $\leq$, $>$,  $\geq$, $==$ and $\neq$.
One operator may be encoded by two or more other operators.
However, the two expressions may have different costs.
We identify some implementation heuristics that help us generate efficient programs.

For example, the HoneyBadgerMPC library provides two comparison protocols: LessThan and Equality. They are used to express $a < b$ and $a == b$ on shared values, and return a secret shared value. Building an MPC compiler requires us to implement other operators in terms of these two.
For example, a naive and intuitive implementation is to encode $a \geq b$ as $(b < a) + (a == b)$. An alternative way is to encode it as $1 - (b < a)$.
Our abstract resource estimator generates one LEQ, one ADD and one EQUAL gate for the first encoding, and one SUB and one LEQ gate for the second encoding.
In the HoneyBadgerMPC resource model, the costs of addition and subtraction are trivial since they require no communication, and the multiplication takes one round and one multicast to finish.
The round complexity of comparison is seven times more than the cost of multiplication \cite{reistad2007secret}, the communication cost is even more expensive. Also, the cost of equality check is higher than the less than operation. 
Thus, we believe that the second encoding is better due to the reduced number of comparison.
This demonstrates how we experiment optimizations guided by our resource estimators.

\begin{table}[b]
   \vspace{-1.5em}
    \small
    \centering
    \caption{Execution time of evaluating $a \geq b$ for 100 times, where $a$ and $b$ are randomly generated number ranging from 1 to 100.}
    \begin{tabular}{ | l | c | }
        \hline
        Encoding $a \geq b$ & Execution Time \\
        \hline
        $(b < a) + (a == b)$ & 0.23s\\
        $1 - (b < a)$ & 0.10s\\
        \hline
    \end{tabular}
    \label{tb:gre}
    \vspace{1.5em}
\end{table}
To verify the above observation, we perform a set of private comparison in a HoneyBadgerMPC program (on
the same machine used in \secref{sec:evaluation}).
Our tests execute 100 times of the greater or equal comparison on two randomly generated numbers. \tbref{tb:gre} compares the running time of the two encodings.

\section{Evaluation}
\label{sec:evaluation}
This section presents three case studies to assess our framework focusing on Harpoon and HIR, optimizing scalar multiplication, and support for indexing arrays with secrets, respectively.
For simplicity, the test program uses plaintext values instead of obtaining them at runtime. 
We conducted our experiments on a machine with 8 Intel Core i7 processors and 16 GB RAM that runs Ubuntu 18.04 LTS. 

\subsection{Case Study 1: Secure Auctions}
\label{sec:auctions}
Recall the discussion of the practical importance of secure auctions in \secref{sec:bkg}.
This experiment implements a second-price auction that is designed to give bidders confidence to bid their best price without overpaying. 
The bidder who submits the highest bid is awarded the item and pays the amount of the second-highest bid.

\figref{fig:2price-harpoon} shows the code snippets in Harpoon, where the elements in arrays \CODE{bidders} and \CODE{bid} denote bidder's identities and their bids.
The implementation uses four variables (\CODE{fst}, \CODE{snd}, \CODE{ifst} and \CODE{isnd}) to store the values of the first and second highest bids and the identities of holders respectively.
As shown in \figref{fig:2price-harpoon}, writing the Harpoon implementation does not require developers to have cryptographic concerns or circuit building mindset.
They can program functionally or imperatively, thanks to the expressiveness of Scala.

As mentioned in \secref{sec:bkg}, our compiler could transform the imperative Harpoon program to a functional style one as \texttt{(bids zip bidders).map(..).reduce(..)}, which yields a circuit of logarithmic depth that allows efficient parallel computation.

\iftoggle{extend}
{
  \figref{fig:spdz-auction} and \figref{fig:hbmpc-auction} show the generated SPDZ and HoneyBadgerMPC program respectively.
  As the loops are unrolled during code generation, we only show the generated program for 3 bidders. 
  For testing and development, the HoneyBadgerMPC program runs in a single-process simulated network, and contains lines of code dealing with network connections and synchronizations. Those are concerns that Harpoon and HIR developers do not have to worry about.  
}{
  We have generated SPDZ and HoneyBadgerMPC programs to realize secure auctions.
  For testing and development, the HoneyBadgerMPC program runs in a  simulated network, and contains lines of code dealing with network connections and synchronizations. 
  The Harpoon and HIR developers need not to have those concerns.
}

\begin{figure}[t]
  \vspace{1em}
  \centering
  \begin{lstlisting}[basicstyle={\scriptsize\ttfamily}]
    @sec var bidders = Array(0, 1 ..., n - 1)
    @sec var bids = Array(b1, b2, ..., bn)
    var ifst = bidders(0)     var isnd = bidders(1)
    var fst = bids(0)         var snd = bids(1)
    if (bids(0) < bids(1)) {
      ifst = bidders(1)   fst = bids(1)
    } else {
      isnd = bidders(1)   snd = bids(1)
    }
    for (i <- 2 until bids.length) {
      if (fst < bids(i)) {
        isnd = ifst         snd = fst
        ifst = bidders(i)   fst = bids(i)
      } else if (snd < bids(i)) {
        isnd = bidders(i)   snd = bids(i)
      }
    }
    (ifst, snd)
  \end{lstlisting}
  \caption{Harpoon code snippet performing a second-price auction, where \CODE{b1, b2, ..., bn} are parameters passed to the method.}
  \label{fig:2price-harpoon}
\end{figure}

\subsection{Case Study 2: Matrix-Vector Product}
Secure matrix-vector multiplication is a core kernel in many real-world applications.
For example, in the area of privacy-preserving machine learning, matrix-vector multiplication is one of the common building blocks of neural networks~\cite{securenn}. 
During the training and inference procedures, it is often the case that multiple parties combine their data where secure matrix-vector multiplication can be used to preserve privacy. 

The case study performs a set of secure matrix-vector multiplication, where one party (the client) has an input matrix, and the other party (the server) has a vector. 
\figref{fig:mat-vec-test} shows the test program that randomly generates a $10 * N$ matrix (where $100 \leq N \leq 500$), and multiplies with a fixed vector $[1, 399, 1, 413, 1, 587, 1, 354, 1, 444]$. 
The test shows the effectiveness of our optimization discussed in \secref{sec:privat-mutl}.

\begin{figure}[t]
    \vspace{-.5em}
    \centering
    \begin{lstlisting}[basicstyle={\footnotesize\ttfamily}]
  val rand = new scala.util.Random
  val start = 1000
  @sec(alice) val m = Array.fill(10)(
                      start + rand.nextInt(start + 1)) 
  val v = Array(1, 399, 1, 413, 1, 587, 1, 354, 1, 444)
  m * v
    \end{lstlisting}
    \caption{Test program of Matrix-Vector Multiplication.}
    \label{fig:mat-vec-test}
    \vspace{-1.5em}
\end{figure}

\iftoggle{extend}
{
The HElib library implements the BGV homomorphic encryption scheme and effectively uses the Smart-Vercauteren ciphertext packing techniques~\cite{Helib}.
}{}
\tbref{tb:matrix-vec} compares the running time of the generated HElib programs with and without optimizations.
As $N$ increases from 100 to 500, the speedups become more observable.
We have provided median of absolute runtime before and after the optimization with 95\% confidence.
\begin{table}[b]
\vspace{-1em}
\small
  \centering
  \caption{Execution time (in seconds) of the HElib programs that perform multiplication of a matrix of $10 * N$ and a vector $[1, 399, 1, 413, 1, 587, 1, 354, 1, 444]$ before and after the optimization.}
  \begin{tabular}{ | l | c | c | c | c | c | }
  \hline
  N & 100 & 200 & 300 & 400 & 500 \\
  \hline
  Before (median) & 7.57 & 21.31 & 55.81 & 135 & 292 \\
  Error (confidence 95\%) & 0.51 & 0.44 & 0.72 & 1.45 & 2.05\\
  \hline
  After (median) & 6.32	& 15.18 & 34.89	& 80	& 162 \\
  Error (confidence 95\%) & 0.05 &	0.34 &	0.23	& 0.34	& 0.49 \\
  \hline
  \end{tabular}
  \label{tb:matrix-vec}
\end{table}
\iftoggle{extend}
{
As our compiler unrolls all the loops, the size of the generated code is not small.
\figref{fig:helib-mul} -- \figref{fig:tfhe-matrxi-vec2} show the generated HElib and TFHE programs that compute matrix-vector multiplication of \CODE{Array(Array(1, 2, 3), Array(4, 5, 6))} and  \CODE{Array(1, 2, 1)}.
}{}

\subsection{Case Study 3: Merge Sort}
\begin{figure}[!ht]
\noindent
\begin{lstlisting}[numbers=left,basicstyle={\scriptsize\ttfamily},escapechar=*, xleftmargin=4.0ex]
val o1 = Owner()    val s = 0
var arr = NumArray(o1, 3, 1, 5, 2)  // input
val e = arr.length
def merge(o: Owner, arr1: NumArray, arr2: NumArray) = {
  var res = NewNumArray(o, arr1.length + arr2.length)
  var i = Num(o, 0)   var j = Num(o, 0)  var k = 0
  while (k < res.length) {   *\label{line:bound}*
    val b1 = i < Num(o, arr1.length)
    val b2 = j < Num(o, arr2.length)
    val p = if (b1.not) arr2(j) else if (b2.not) arr1(i)
            else if (arr1(i) <= arr2(j)) arr1(i) else arr2(j)
    res = res.update(k, p)
    // updating arr1 index
    i = if (b1.not) i else if (b2.not) i + Num(o, 1) else
        if (p == arr1(i)) i + Num(o, 1) else i
    // updating arr2 index
    j = if (b1.not) j + Num(o, 1) else if (b2.not) j
        else if (p == arr2(j)) j + Num(o, 1) else j
    k = k + 1
  }
  res
}
val r = recFuel(10);  *\label{line:fuel}*
val mergesort = r.rec[NumArray, Owner, Int, Int] { *\label{line:merge}*
                f => (a, o, i, j) => {
  val mid = (j - i) / 2
  if (mid == 0 || i >= j){ a }  *\label{line:ifs}*
  else {
    val left = a.slice(i, mid)  val right = a.slice(mid, j)
    merge(o, f(left, o, 0, left.length),               
             f(right, o, 0, right.length))}  *\label{line:ife}*
  }
}
val res = mergesort(arr, o1, s, e)
output(res.eval(o1))
\end{lstlisting}
\caption{MergeSort implemented in HIR.}
\label{fig:merge-sort-hir}
\vspace{-1em}
\end{figure}
MergeSort is a key computation component of various Secure MPC applications. 
For example, when multiple parties exchange messages anonymously, both the content and the metadata (\eg, the length of the message) need to be protected.
Secure sort is one of the core kernels used for such anonymous communications~\cite{mcmix}. 

This case-study implements MergeSort in HIR,  
as it exposes the language features needed in writing secure computation. The implementation involves array indexing and conditional executions. Notably, an array lookup on a private index is not supported by most programming languages~\cite{hastings2019sok}.

MergeSort recursively divides an input array into two halves and then merges the two sorted halves. Our implementation is shown in \figref{fig:merge-sort-hir}.
In the function \CODE{mergesort}, the variable \CODE{r} (line \ref{line:fuel}) stores a recursion object initialized with the bound 10. The expression \CODE{r.rec} (line \ref{line:merge}) is the construct for defining a bounded recursive function call. This allows one to explicitly specify the bound of the defining recursive function. The \CODE{NumArray} is the type for arrays that allow private indexing. The two parameters \CODE{i} and \CODE{j} are plaintext, which is important for unrolling the recursive function at compile time.
The function \CODE{slice(i, j)} returns a subarray from the \CODE{i}th element until the \CODE{j}th element, where \CODE{i} and \CODE{j} are plaintext integers. The if-statement (lines \ref{line:ifs} to \ref{line:ife}) is the standard one as its condition depends on a plain text value.
The function \CODE{merge} is used for merging two halves.
All the if-constructs appearing in this function are oblivious as their conditions depend on ciphertext values. The loop (line \ref{line:bound}) is bounded as the length of an array is known at compile time.
\iftoggle{extend}
{
\figref{fig:tfhe-merge1} -- \figref{fig:tfhe-merge4} show the generated TFHE program, where loops and recursions are all unrolled.
}{}
\section{Related Work}
\label{sec:relatedwork}
There have been many MPC frameworks proposed in recent years and several of them are already integrated into HACCLE. We list the prominent MPC frameworks as follows.

SCALE-MAMBA \cite{scale-mamba} is an existing MPC framework that is closest to HACCLE.  
We utilize it as one of our cryptographic backends to implement secret sharing and FHE based protocols. 
It is a combination of a compiler and a run-time environment where optimizations can be performed at a lower level. 
Compared with SCALE-MAMBA, HACCLE provides staging driven by type systems, estimates resource consumption, and focuses on optimization at a higher level.

HoneybadgerMPC~\cite{lu2019honeybadgermpc} is another backend of HACCLE that supports secret-sharing based protocols. 
The uniqueness of HoneybadgerMPC is the combination of a robust online phase and an optimal non-robust offline phase. 
It provides fairness guarantees even in the asynchronous network setting and also preserves efficiency to make MPC programs practical to run.

As privacy preserving machine learning becomes more and more popular, many frameworks have been developed specifically for this use case, such as ABY \cite{aby}, ABY3~\cite{aby3}, CHET \cite{chet}, EzPC \cite{ezpc}, CrypTFlow \cite{kumar2019cryptflow} and SecureNN \cite{securenn}. 
These frameworks are highly optimized for machine learning and are designed for two-party or three-party settings. 
We choose not to include them due to our desire to support an arbitrary number of parties.
There are also many other MPC frameworks such as Viff \cite{vifflib}, Jiff \cite{jiff}, MPyC \cite{mpyc} and PICCO \cite{zhang2013picco}. 
Theoretically, any framework can be embedded as a backend in HACCLE even though not all of them are integrated at the moment. 
\vspace{-1ex}

\section{Conclusion}
\label{sec:conclusion}

Secure MPC-based applications play a crucial role in solving many important practical problems such as in high-value procurement.
But developing performant MPC-based applications from scratch is a notoriously difficult task as it requires expertise ranging from cryptography to circuit optimization.
Therefore software developers need a compiler toolchain for developing MPC-based applications.
As a solution to this problem, we have introduced the HACCLE toolchain, a multi-stage compiler for optimized circuit generation.
We believe that the HACCLE toolchain offers a compelling approach to the design and implementation of Secure MPC applications, using metaprogramming techniques.

\begin{acks}
We thank the anonymous reviewers for their helpful suggestions and comments.
This research is based upon work supported by the Office of the Director of National Intelligence (ODNI), Intelligence Advanced Research Projects Activity (IARPA), contract \#2019-19020700004. 
The views and conclusions contained herein are those of the authors and should not be interpreted as necessarily representing the official policies, either expressed or implied, of ODNI, IARPA, or the U.S. Government. 
The U.S. Government is authorized to reproduce and distribute reprints for governmental purposes notwithstanding any copyright annotation therein.
\end{acks}

\bibliographystyle{ACM-Reference-Format}
\balance
\bibliography{paper}

\iftoggle{extend}{
  \clearpage
  \appendix
  \section{Appendix}
\label{sec:appA}

\subsection{Harpoon Syntax}
\label{sec:syntax-harpoon}
\figref{fig:harpoon-syntax} shows the abstract syntax of Harpoon. We
note that the concrete syntax used in our example Harpoon programs is
slightly different-- our abstract syntax does not include code
comments, for example. Harpoon is parameterized over sets of unique
identifiers for variables, functions, and owners: $\mathbb{I}_v$,
$\mathbb{I}_f$, and $\mathbb{I}_o$, respectively. Harpoon programs
begin with a header specifying the bit-width of integer values: while
this does not effect the semantics or typing of Harpoon programs, this
parameter is utilized by subsequent stages of the pipeline to generate
the final computation. This header is followed by a collection of
function definitions. One of these functions is tagged with
@\keyw{EntryPoint} to designate it as the entry point for computation,
akin to the \lstinline|main| function of a C program or the
\lstinline|main| method of a Java class. The ability to define new
functions is a key feature of Harpoon; this is the mechanism by which
we can add new abstractions to the language.

A function is defined by a list of parameters and their types, the
return type of the function, an upper bound on the number of function
calls, and a function body. Harpoon has three categories of types: an
\emph{atomic type} is either an integer, double, or boolean; a
\emph{base type} is either an atomic type or an array of atomic types;
and \emph{compound types} is either a base type with an optional
\emph{security annotation} or a pair of compound types. Security
annotations are used to mark high-security or private data. The
annotations for any high-security inputs or outputs of the entry point
functions include an explicit \emph{owner}; while Harpoon's type
system does not make use of fine-grained ownership information (see
\autoref{sec:HarpoonTyping} for more details), these details will be
passed down through the pipeline. In particular, the HACCLE pipeline
uses these ownership annotations to generate the glue code to retrieve
and assemble inputs arguments from concrete network nodes and to share
the results of the computation with concrete nodes.  Fine-grained
ownership information can also used to enforce backend- and network-
specific information flow policies.

The HACCLE toolchain allows parties to internally represent their
private data using arbitrary Scala objects. At present, however, the
toolchain HACCLE requires the arguments to and results from a
computation to be one of Harpoon's base types, \ie, an atomic type or
an array of atomic types. Thus, each logical owner of an argument to
the entry point function is responsible for encoding their data as a
Harpoon type; the compiler generates the glue code routing these
inputs into the computation. The toolchain also generates code that
routes each component of the result to the owner specified by its
annotation; each recipient of such a value is similarly responsible
for decoding it into their internal representation of the data. We
plan to automate these sorts of encoding and decoding steps in the
future.

Function bodies consist of a sequence of statements, followed by an
expression specifying the result value to be returned to the caller of
the function. We restrict function bodies to a single return statement
for simplicity; we may relax this restriction in future iterations of
Harpoon. Finally, each function definition is annotated with an
expression placing an upper bound on the number of recursive
calls. This expression can reference the parameters of the function,
allowing this bound to vary according to the context in which a
function is called. 

The syntax of
Harpoon expressions is largely straightforward and includes constants,
variables, function calls, and built-in operations on atomic values,
arrays, and pairs. Harpoon does not have ad-hoc polymorphism,
\ie, operator overloading, and thus has distinct operators for
integers and doubles. We denote whether a built-in operation has
integer, double or boolean arguments via ${\mathbb{I}}$,
${\mathbb{D}}$, and ${\mathbb{B}}$ subscripts. The set of operations
is designed to be in one-to-one correspondence with nodes in HIR,
hence the absence of certain operators for doubles (\eg, modulo). We
plan to add new operators to this set as HIR evolves.
\begin{figure*}
  \centering
\[
  \begin{array}{l l l l}
    \text{Variable Identifiers} & x & \in & \mathbb{I}_v \\
    \text{Function Identifiers} & f & \in & \mathbb{I}_f \\
    \text{Owner Labels} & o & \in & \mathbb{I}_o \\
    \text{Program} & P & ::= & H\ D_{1} \ldots D_{n} ~ D_F \\
    \text{Header} & H & ::= & @\keyw{IntWidth}(c) \\
    \text{Function Definition} & D & ::= & \keyw{def}\ f(x_{1}: \tau_{1}, \ldots, x_{n}: \tau_{n}): \tau\ B = \{S;e\} \\
    \text{Entry Point Definition} & D_F & ::= & @\keyw{EntryPoint}\
    \keyw{def}\ f(x_{1}: \beta_{1}\ {O_1}, \ldots, x_{n}: \beta_{n}\ {O_n}): \tau^{O}~B = \{S;e\} \\\\
    \text{Atomic Type} & \alpha & ::= & \keyw{Boolean} \mid \keyw{Int} \mid \keyw{Double} \\
    \text{Base Type} & \beta & ::= & \alpha \mid \keyw{Array}[\alpha] \\
    \text{Compound Type} & \tau & ::= & \beta~M \mid \tau \times \tau \\
    \text{Wire Type} & \tau^{O} & ::= & \beta~O \mid \tau^{O} \times \tau^{O} \\
    \text{Security Annotation} & M & ::= & \emp \mid @\keyw{Sec} \\
    \text{Owner Annotation} & O & ::= & \emp \mid\ @\keyw{Sec}(o) \\
    \text{Bound Annotation} & B & ::= &\ @\keyw{Bound}(e) \\\\
    \text{Statement} & S & ::= & \keyw{skip} ~ \mid ~ S;S ~ \mid ~ \keyw{var}\ x = e ~ \mid ~ x = e \mid x(e) = e ~ \mid ~ \keyw{if}(e)\{S\}\keyw{else}\{S\} \\
    &&\mid& \keyw{for}(x \leftarrow e\ B\ \keyw{until}\ e\ B)\{S\} ~\mid~  \keyw{while}(e\ B)\{S\} \\
    \text{Expression} & e & ::= & c \mid x \mid x(e) \mid  \odot(e_{1}, \ldots, e_{n}) \mid  f(e_{1}, \ldots, e_{n}) \mid \keyw{Array}.\keyw{fill}(e, e) \mid e.\keyw{length} \mid e.\keyw{slice}(e, e) \\
    &&\mid& (e, e) \mid e.\keyw{fst} \mid e.\keyw{snd} \\
    \text{Primitive Operation} & \odot & ::= & \neg \mid \land \mid \lor \mid \oplus \mid +_{\mathbb{I}} \mid -_{\mathbb{I}} \mid \times_{\mathbb{I}} \mid \div_{\mathbb{I}} \mid \bmod \mid \gg \mid \ll \mid \&\& \mid || \mid +_{\mathbb{D}} \mid -_{\mathbb{D}} \mid \times_{\mathbb{D}} \mid \div_{\mathbb{D}} \\
    &&\mid& <_{\mathbb{I}} \mid \le_{\mathbb{I}} \mid >_{\mathbb{I}} \mid \ge_{\mathbb{I}} \mid <_{\mathbb{D}} \mid \le_{\mathbb{D}} \mid >_{\mathbb{D}} \mid \ge_{\mathbb{D}} \mid  =_{\mathbb{B}} \mid =_{\mathbb{I}} \mid =_{\mathbb{D}} \mid \mathrm{toDouble} \mid \mathrm{toInt} \\\\
    \text{Integer Constants} & i & \in & \mathbb{Z} \\
    \text{Double Constants} & d & \in & \mathbb{D} \\
    \text{Constant} & c & ::= & \keyw{true} \mid \keyw{false} \mid i \mid d \\
  \end{array}
\]
  \caption{The abstract syntax of Harpoon}
  \label{fig:harpoon-syntax}
\end{figure*}
\subsection{Harpoon Type System}
\label{sec:HarpoonTyping}
Harpoon is a \emph{security-typed language}~\cite{zdancewic2003} whose
type system is designed to statically prevent private data leaking via
public channels. Guaranteeing that computations involving low-security
variables are not influenced by high-security data at the Harpoon
level allows later stages in the HACCLE pipeline to safely compute
low-security information in the clear in order to produce more
performant secure computations (see \secref{sec:haccle} for more
details). This stage of the pipeline does not utilize the fine-grained
ownership information included in the definition of the entry point
function; thus Harpoon's type system only identifies data as either
high- or low-security.

\figref{fig:harpoon-type} presents the types used by Harpoon's type
system. The type system\footnote{We overload the symbol $\tau$ to
  represent to both labeled and annotated composite types; for the
  rest of this section it refers to the former.} adopts a more
standard notation for security labels instead of the \lstinline|Sec|
labels used by our source syntax. Formally, security labels now range
over elements of a security lattice $\mathcal{L}$, which contains two
elements, $\shi$ and $\slo$.  This lattice is equipped with the
conventional order relation, $\sle$, as well as join and meet
operations, $\scup$ and $\scap$ respectively. It is straightforward to
convert compound and wire types to labeled types: the annotations
$@\keyw{Sec}$ and $@\keyw{Sec}(o)$ used function and entry point
definitions both map to the high-security label $\shi$, whereas an
unannotated type is given the low-security level label $\slo$.

The security lattice also induces a subtyping relation on types which
captures the fact that it is safe to use low-security data when
high-security data is required, but not vice-versa. The subtyping
relation is of the form
\[ \tau_{1} <: \tau_{2} \] which is read as ``type $\tau_{1}$ is a
subtype of type $\tau_{2}$''. Harpoon's subtyping rules are shown in
\figref{fig:harpoon-subtyp}. Note that the subtyping relation is both
reflexive and transitive.

Our typing relations rely on two contexts: one mapping function
identifiers to their signatures and another mapping variables to a
labeled compound type and access annotation. Access annotations are
used to ensure that low-security variables are not updated inside of a
high-security block, e.g. one of the branches of an \keyw{if} statement
whose condition depends on some high-security information. To enforce
this restriction, some of our typing rules rely on an auxiliary
function $\keyw{restrict}(\Gamma, l)$ which builds a typing context
where every variable of security level lower than $l$ is marked as
read-only. More formally, every entry $x : \tau\ w$, in $\Gamma$, is
mapped to $x : \tau\ \keyw{RO}$ if $\neg l \sle \keyw{lb}(\tau)$, where
$\keyw{lb}(\tau)$ takes the lower-bound of the security labels in
$\tau$. $\keyw{lb}$ is defined naturally as follows:
\[
\begin{array}{lcl}
  \keyw{lb}(\beta^{l}) &=& l \\
  \keyw{lb}(\tau_{1} \times \tau_{2}) &=& \keyw{lb}(\tau_{1}) \scap \keyw{lb}(\tau_{2}) \\
\end{array}
\]

\begin{figure}
  \centering
  \begin{mathpar}

    \inferrule*[right=SLab]{
      l \sle l'
    } {
      \beta^{l} <: \beta^{l'}
    }

    \inferrule*[right=SProd]{
      \tau_1 <: \tau_1' \\
      \tau_2 <: \tau_2'
    } {
      \tau_1 \times \tau_2 <: \tau_1' \times \tau_2'
    }

  \end{mathpar}

  \caption{Subtyping rules}
  \label{fig:harpoon-subtyp}
\end{figure}

\begin{figure}
  \centering
  \[
  \begin{array}{l l l l}
    \text{Labeled Compound Type} & \tau &::=& \beta^{l} \mid \tau \times \tau \\
    \text{Atomic Operation Type} & U &::=& \alpha_{1} \times \cdots \times \alpha_{n} \to \alpha \\
    \text{Function Type} & T &::=& \tau_{1} \times \cdots \times \tau_{n} \to \tau \\
    \text{Security Label} & l &::=& \shi \mid \slo \\\\
    \text{Variable Typing} Context & \Gamma &::=& \emp \mid \Gamma, x: \tau\ w \\
    \text{Access Annotation} & w &::=& \emp \mid \keyw{RO} \\
    \text{Function Typing Context} & \Psi &::=& \emp \mid \Psi, f: T \\
  \end{array}
  \]
  \caption{Syntax of Harpoon Types.}
  \label{fig:harpoon-type}
\end{figure}

\subsubsection{Typing Rules}
The rules defining the typing relation over Harpoon expressions,
statements, and programs are given in
Figures~\ref{fig:harpoon-exp-typ}, \ref{fig:harpoon-stm-typ}, and
\ref{fig:harpoon-prg-typ}, respectively. The judgment for expressions
has the form
\[ \etp\Psi\Gamma{e}{T} \] which is read as ``expression $e$ has type
$T$ under function typing context $\Psi$ and variable typing context
$\Gamma$''. These rules include the subsumption rule \textsc{TSub}
which uses the subtyping relation to promote from low- to
high-security expressions. The \textsc{TOp} rule uses a function that
maps atomic operations to their signatures, analogously to the
$\den{\cdot}$ function used in \textsc{EOp}. This rule also enables a
kind of security polymorphism by using the meet operation of the
security lattice to label the result type with the maximum security
label attached to its arguments. The typing rule for function calls,
\textsc{TCall}, relies on the function typing context $\Psi$ to get
the function's signature, as expected.

Harpoon's type system requires that the length of every arrays be
public, a design decision reflected in the typing rules for array
operations. Arrays with secret array length cannot be compiled into input-independent circuits\footnote{We do note that secret array lengths can be partially supported by providing a publicly-known {\em upper bound} on array length---the circuits that manipulate this array will use this upper bound.}. This choice
is obvious in the \textsc{TLen} rule, which always returns a
low-security integer. The \textsc{TFill} rule forces a low-security
expression be used to determine the length of the newly created array;
the \textsc{TSlice} rule similarly requires the indices used for
slicing to be low-security.

\begin{figure}[t]
  \centering
  \begin{mathpar}

    \inferrule*[right=TConst.Bool]{
      b \in \{\keyw{true}, \keyw{false}\}
    } {
      \etp\Psi\Gamma{b}{\keyw{Boolean}^{\slo}}
    }

    \inferrule*[right=TConst.Int]{
    } {
      \etp\Psi\Gamma{n}{\keyw{Int}^{\slo}}
    }

    \inferrule*[right=TConst.Double]{
    } {
      \etp\Psi\Gamma{d}{\keyw{Double}^{\slo}}
    }

    \inferrule*[right=TVar]{
      \Gamma(x) = T\ w
    } {
      \etp\Psi\Gamma{x}{T}
    }

    \inferrule*[right=TSub]{
      \etp\Psi\Gamma{e}{T} \\
      T <: T'
    } {
      \etp\Psi\Gamma{e}{T'}
    }

    \inferrule*[right=TOp]{
      \odot : \alpha_1 \times \cdots \times \alpha_n \to \alpha \\
      \overline{\etp\Psi\Gamma{e_i}{\alpha_i^{l_i}}} \\
    } {
      \etp\Psi\Gamma{\odot(e_1, \ldots, e_n)}{\alpha^{\scup l_i}}
    }

    \inferrule*[right=TCall]{
      \Psi(f) = \tau_{1} \times \cdots \times \tau_{n} \to \tau \\
      \overline{\etp\Psi\Gamma{e_i}{\tau_i}}
    } {
      \etp\Psi\Gamma{e(e_1, \ldots, e_n)}{\tau}
    }

    \inferrule*[right=TFill]{
      \etp\Psi\Gamma{e_1}{\keyw{Int}^{\slo}} \\
      \etp\Psi\Gamma{e_2}{\alpha^{l}}
    } {
      \etp\Psi\Gamma{\keyw{Array}.\keyw{fill}(e_1, e_2)}{\keyw{Array}[\alpha]^{l}}
    }

    \inferrule*[right=TSel]{
      \etp\Psi\Gamma{e}{\keyw{Array}[\alpha]^{l}} \\
      \etp\Psi\Gamma{e'}{\keyw{Int}^{l'}} \\
      l' \sle l
    } {
      \etp\Psi\Gamma{e(e')}{\alpha^{l}}
    }

    \inferrule*[right=TLen]{
      \etp\Psi\Gamma{e}{\keyw{Array}[\alpha]^{l}} \\
    } {
      \etp\Psi\Gamma{e.\keyw{length}}{\keyw{Int}^{\slo}}
    }

    \inferrule*[right=TSlice]{
      \etp\Psi\Gamma{e}{\keyw{Array}[\alpha]^{l}} \\
      \etp\Psi\Gamma{e_1}{\keyw{Int}^{\slo}} \\
      \etp\Psi\Gamma{e_2}{\keyw{Int}^{\slo}}
    } {
      \etp\Psi\Gamma{e.\keyw{slice}(e_1, e_2)}{\keyw{Array}[\alpha]^{l}}
    }

    \inferrule*[right=TProd]{
      \etp\Psi\Gamma{e_1}{\tau_1} \\
      \etp\Psi\Gamma{e_2}{\tau_2}
    } {
      \etp\Psi\Gamma{(e_1, e_2)}{\tau_1 \times \tau_2}
    }

    \inferrule*[right=TFst]{
      \etp\Psi\Gamma{e}{\tau_1 \times \tau_2}
    } {
      \etp\Psi\Gamma{e.\keyw{fst}}{\tau_1}
    }

    \inferrule*[right=TSnd]{
      \etp\Psi\Gamma{e}{\tau_1 \times \tau_2}
    } {
      \etp\Psi\Gamma{e.\keyw{snd}}{\tau_2}
    }

  \end{mathpar}

\caption{Typing rules for expressions}
\label{fig:harpoon-exp-typ}
\end{figure}

The typing judgment for statements is similar and has the form
\[ \stp\Psi\Gamma{l_{pc}}{S}{\Gamma'} \] which reads ``statement
$S$ is well typed under function typing context $\Psi$, variable
typing context $\Gamma$, and program counter label $l_{pc}$,
producing new context $\Gamma'$''. The program counter label
$l_{pc}$ is used to monitor the security level of the information
that can be learned by knowing the program has reached the current
statement. To see how this label is used, consider the following
Harpoon code snippet which uses a low-security value $l$ and
a high-security value $h$:
\begin{align*}
& l = 0; \\
& \keyw{if}~ (h=_\mathbb{I} 1) ~\keyw{then}~ \{l := 1\} ~\keyw{else}~ \{\keyw{skip}\}
\end{align*}
The update to the value of $l$ in the $\keyw{then}$ branch leaks
information about the value of $h$ and should clearly be
disallowed. More generally, if a high-security value is used in the
test expression, we should not leak any information about which branch
was taken by making an assignment to a low-security variables declared
outside of the block. The typing rule for conditional statements,
\textsc{TIf} disallows such side channels via the assumption
$l' = l \scup l_{pc}$, which elevates the security of program
counter label used to the branches, $l'$ to the upper bound of the
current program counter label $l_{pc}$ and the security level of
the test expression $l$. The rule further forces all variables with
a lower security level than $l$ to be read-only via the previously
detailed $\keyw{restrict}$ function. Combined with the restriction that
read-only variables and arrays cannot be assigned or updated imposed
by \textsc{TAss} and \textsc{TUpd}, this restricted context ensures
that information about high-security branch conditions cannot be
leaked via low-security variables. The rules for loops make similar
use of the program counter label to prohibit programs like:
\begin{align*}
& l = 0; \\
& \keyw{while}~ (h<_\mathbb{I}10~@\keyw{Bound}(10)) \{l = l +_\mathbb{I} 1\}
\end{align*}
The typing rules for loops also require that the upper bounds on loop
iterations be low-security, a requirement needed to ensure that any
loop unrolling done by later stages in the pipeline does not leak
high-security information.

\begin{figure*}
  \centering
  \begin{mathpar}

    \inferrule*[right=TSkip]{
    } {
      \stp\Psi\Gamma{l_{pc}}{\keyw{skip}}{\Gamma}
    }

    \inferrule*[right=TSeq]{
      \stp\Psi\Gamma{l_{pc}}{S_1}\Gamma' \\
      \stp\Psi{\Gamma'}{l_{pc}}{S_2}{\Gamma''}
    } {
      \stp\Psi\Gamma{l_{pc}}{S_1; S_2}{\Gamma''}
    }

    \inferrule*[right=TNew]{
      \etp\Psi\Gamma{e}{\tau}
    } {
      \stp\Psi\Gamma{l_{pc}}{\keyw{var}\ x = e}{\Gamma, x : \tau}
    }

    \inferrule*[right=TAss]{
      \Gamma(x) = \tau\ w \\
      \etp\Psi\Gamma{e}{\tau} \\
      w \ne \keyw{RO}
    } {
      \stp\Psi\Gamma{l_{pc}}{x = e}{\Gamma}
    }

    \inferrule*[right=TUpd]{
      \Gamma(x) = \keyw{Array}[\alpha]^{l}\ w \\
      \etp\Psi\Gamma{e_1}{\keyw{Int}^{l'}} \\
      l' \sle l \\
      \etp\Psi\Gamma{e_2}{\alpha^{l}} \\
      w \ne \keyw{RO}
    } {
      \stp\Psi\Gamma{l_{pc}}{x(e_1) = e_2}{\Gamma}
    }

    \inferrule*[right=TFor]{
      \etp\Psi\Gamma{e_1}{\keyw{Int}^{l_1}} \\
      \etp\Psi\Gamma{e_2}{\keyw{Int}^{l_2}} \\
      \etp\Psi\Gamma{e_1'}{\keyw{Int}^{\slo}} \\
      \etp\Psi\Gamma{e_2'}{\keyw{Int}^{\slo}} \\\\
      \stp\Psi{\Gamma, x: \keyw{Int}^{\slo}\ \keyw{RO}}{l_{pc}}{S}{\Gamma'}
    } {
      \stp\Psi\Gamma{l_{pc}}{\keyw{for}(x \leftarrow e_1\ @\keyw{Bound}(e_1')\ \keyw{until}\ e_2\ @\keyw{Bound}(e_2'))\{S\}}{\Gamma}
    }

    \inferrule*[right=TWhile]{
      \etp\Psi\Gamma{e}{\keyw{Boolean}^{l}} \\
      \etp\Psi\Gamma{e'}{\keyw{Int}^{\slo}} \\
      \stp\Psi\Gamma{l_{pc}}{S}{\Gamma'}
    } {
      \stp\Psi\Gamma{l_{pc}}{\keyw{while}(e\ @\keyw{Bound}(e'))\{S\}}{\Gamma}
    }

    \inferrule*[right=TIf]{
      \etp\Psi\Gamma{e}{\keyw{Boolean}^{l}} \\
      l' = l \scup l_{pc} \\
      \Gamma' = \keyw{restrict}(\Gamma, l') \\
      \stp\Psi{\Gamma'}{l'}{S_1}{\Gamma_1} \\
      \stp\Psi{\Gamma'}{l'}{S_2}{\Gamma_2}
    } {
      \stp\Psi\Gamma{l_{pc}}{\keyw{if}(e)\{S_1\}\keyw{else}\{S_2\}}{\Gamma}
    }

  \end{mathpar}

\caption{Typing rules for statements}
\label{fig:harpoon-stm-typ}
\end{figure*}

The rules for typing Harpoon function definitions and programs are
given in \figref{fig:harpoon-prg-typ}. The conversion from compound
and wire types to labeled types is denoted as $\lfloor \tau \rfloor$
in these rules. Typing Harpoon functions is straightforward: a
function definition $D$ is well-formed under function typing context
$\Psi$ if its signature agrees with the one in $\Psi$, its body and
return expression are well-typed, and its bound only depends on public
information. Similar to loop bounds, this last restriction ensures
that any function inlining done by later stages of the pipeline does
not leak any information. Finally, a Harpoon program is well formed if
each of its functions is well formed.

\begin{figure*}
  \centering
  \begin{mathpar}

    \inferrule*[right=TDef]{
      \Psi(f) = {\lfloor\tau_{1} \times \cdots \times \tau_{n} \to \tau\rfloor} \\
      \Gamma = x_{1} : \lfloor\tau_{1}\rfloor\ \keyw{RO}, \ldots, x_{n} : \lfloor\tau_{n}\rfloor\ \keyw{RO} \\
      \etp{\Psi}{\Gamma}{e'}{\keyw{Int}^{\slo}} \\
      \stp{\Psi}{\Gamma}{\slo}{S}{\Gamma'} \\
      \etp{\Psi}{\Gamma'}{e}{\lfloor\tau\rfloor}
    } {
      \dtp{\Psi}{\keyw{def}\ f(x_{1}: \tau_{1}, \ldots, x_{n}: \tau_{n}): \tau\ @\keyw{Bound}(e') = \{S;e\}}
    }

    \inferrule*[right=TEDef]{
      \Psi(f) = {\lfloor\beta_{1}\ {O_1} \times \cdots \times
        \beta_{n}\ {O_n} \to \tau^{O}\rfloor} \\
      \Gamma = x_{1} : \lfloor\beta_{1}\ {O_1}\rfloor\ \keyw{RO},
      \ldots, x_{n} : \lfloor\beta_{n}\ {O_n}\rfloor\ \keyw{RO} \\
      \etp{\Psi}{\Gamma}{e'}{\keyw{Int}^{\slo}} \\
      \stp{\Psi}{\Gamma}{\slo}{S}{\Gamma'} \\
      \etp{\Psi}{\Gamma'}{e}{\lfloor\tau^{O}\rfloor}
    } {
      \dtp{\Psi}
      {~@\keyw{EntryPoint}\ \keyw{def}\ f_e(x_{1}: \beta_{1}\ {O_1},
        \ldots, x_{n}: \beta_{n}\ {O_n}):
        \tau^{O}\ @\keyw{Bound}(e') = \{S;e\}}
    }

    \inferrule*[right=TProg]{
      \Psi = f_1 \mapsto \lfloor\overline{\tau_1} \to \tau_1\rfloor, \ldots, f_n
      \mapsto \lfloor\overline{\tau_n} \to \tau_n\rfloor,
      f_e \mapsto \lfloor\overline{\beta\ {O}} \to \tau^{O}\rfloor\\
      \text{No duplicate names in }\Psi \\
      \overline{\dtp{\Psi}{\keyw{def}\ f(\overline{x: \tau}): \tau\ B =
          \{S;e\}}} \\
      \dtp{\Psi}{~@\keyw{EntryPoint}\ \keyw{def}\ f_e(\overline{x: \beta\ {O}}):
        \tau^{O}~B = \{S;e\}}
    } {
      \ptp{H~\overline{\keyw{def}\ f(\overline{x: \tau}): \tau\ B = \{S;e\}}
        ~@\keyw{EntryPoint}\ \keyw{def}\ f_e(\overline{x: \beta\ {O}}):
        \tau^{O}~B = \{S;e\}
      }
    }

  \end{mathpar}

\caption{Typing rules for definitions and programs}
\label{fig:harpoon-prg-typ}
\end{figure*}
\subsection{HIR Syntax}
\label{sec:syntax-hir}
\figref{fig:syntax} shows the syntax of the HIR. Contributors are modeled by set of integers. Inspired by WYSTERIA \cite{rastogi2014wysteria}, the HIR allows contributors to be values, which can be constructed from singleton sets $\{n\}$ and set unions $o_1 \cup o_2$. Atomic types denote plaintext values. Types $[\alpha]_n$ denote array values, where $n$ is the array length. Composite types  denote ciphertext values, and have two forms. The first form $(\alpha, o)$ consists of two parts: an atomic type such as \keyw{int}, and an expression that denotes the set of contributors who provide the data, called providers. For example, an expression with type $(\keyw{int}, \{1\})$ means the singleton set of contributors $\{1\}$ provides an integer value. The two parts of this composite type act largely independently. Any values in the first part may be accessed or observed in any way described in the second part. 
The second form $\keyw{sh}_{(o_1, n)}(\alpha, o_2)$ denotes shared values, and means that a value with type $\alpha$ is shared among the set of contributors (called players) $o_1$ with threshold $n$. And once the value is combined, it may be observed by the set of contributors $o_2$ (called observers). Note nobody is allowed to observe a shared value until it gets combined. For example, a shared type $\keyw{sh}_{(\{1, 2\}, 1)}(5, \{1\})$ means that the integer 5 is shared between two players 1 and 2 with threshold 1. And once the value is combined, only player 1 can observe it. Similarly to the first form of composite types, the values in the first part of the pair are computed independently with who provides it, who holds a share of it once it is shared, and who can observe it once the shares are combined.

Expressions $\Expr$ include the integer constants $n$, bool constants \nonterm{true} and \nonterm{false} \footnote{The implementation has type \texttt{Bit}, which is isomorphic to \keyw{bool}. And an array of \texttt{Bit} is isomorphic to an array of \keyw{bool}.}, variables $x$, array literals $\{\overline{\Expr}\}_\nonterm{n}$ (where $n$ is the size), array reads $x[\Expr]$, array update, array slice, unary operations $\Diamond_u \Expr$, binary operations $\Expr_1 \Diamond_b \Expr_2$, and conditional expressions $\Expr?\Expr_1\!\!:\!\!\Expr_2$. The implementation contains more operations that can be encoded by one or more operations listed in the grammar, thus are omitted for simplicity. For example, the NOR operation may be encoded by $\&\&$ and $\neg$.  The expression $\keyw{share}(e, o_1, o_2, n)$ shares the value of $e$ with the set of players $o_2$ with threshold $n$. And the value may be observed by $o_2$ once it is combined. The expression $\keyw{reshare}(e, o)$ re-shares the value of $e$ with the set of players $o$. The expression $\keyw{combine}(e, o)$ combines the shared value of $e$ from the set of players $o$. The expression $\keyw{eval}(\Expr, o)$ reveals the value of $\Expr$ to the set of players $o$. 

Statements $\Stmt$ comprise of variable declarations, assignments, and a sequence of statements.  
\begin{figure*}[t]
\centering
\[
\begin{array}{l l l l l}
\text{Contributors} & \nonterm{o} & ::= & \{\nonterm{n}\} ~|~ \nonterm{o}_1\cup \nonterm{o}_2 \\
\text{Atomic Type} & \alpha & ::= & \keyw{int} ~|~ \keyw{bool} ~|~ \keyw{float}  ~|~ [\alpha]_\nonterm{n} \\
\text{Composite Type} & \beta & ::= & (\alpha,\nonterm{o})  ~|~ \keyw{sh}_{(\nonterm{o}_1, \nonterm{n})}(\alpha,\nonterm{o}_2) \\
\text{Type} & \nonterm{t}  & ::= & \alpha  ~|~ \beta  ~|~ \gamma \\
\text{Expression} & \nonterm{\Expr} & ::= & \nonterm{n}  ~|~ \nonterm{true}  ~|~ \nonterm{false}  ~|~ \nonterm{x}  ~|~ \{\overline{\Expr}\}_\nonterm{n}  ~|~ \nonterm{x}[\nonterm{\Expr}]  ~|~ \nonterm{x}.\keyw{upd}(\Expr_1,\Expr_2)  ~|~  \nonterm{x}.\keyw{slc}(n_1,n_2)  ~|~ \keyw{share}(\Expr,\nonterm{o}_1,\nonterm{o}_2,\nonterm{n})   \\
& & | &  \keyw{reshare}(\Expr,\nonterm{o}) ~ | ~ \keyw{combine}(\Expr,\nonterm{o}_1,\nonterm{o}_2)  ~|~  \keyw{eval}(\Expr,\nonterm{o})  ~|~ \Diamond_u\nonterm{\Expr}  ~|~ \nonterm{\Expr}_1\Diamond_b\nonterm{\Expr}_2  ~| ~ \nonterm{\Expr}?\nonterm{\Expr}_1:\nonterm{\Expr}_2 \\
\text{Statement} & \nonterm{\Stmt} & ::= & \keyw{skip};  ~|~ \keyw{val} \nonterm{x}:\nonterm{t}:=\nonterm{\Expr};  ~|~ \keyw{val} \nonterm{x} := \Expr; ~ | ~  \nonterm{\Stmt}_1\nonterm{\Stmt}_2 \\
& \nonterm{$\Diamond_u$} & ::= &  -  ~|~ \neg \\
& \nonterm{$\Diamond_b$} & ::= &  +  ~|~ -  ~|~ \ast  ~|~ /  ~|~ \%  ~|~ \ll \| \gg  ~|~ \leq  ~|~ <  ~|~ \geq \| > \| = \\ 
& & | &  \&\&  ~|~ ||  ~|~ ++ %
\end{array}
\]
\caption{The abstract syntax of the intermediate language}
\label{fig:syntax}
\end{figure*}

\subsection{HIR Type System}
\label{sec:hir-typing-rules}
The goal of the type checking of the first part of a composite type is to ensure that the static type of each expression is a supertype of the actual, run-time type of every value it might produce. The goal of the type checking of the rest of a type construct is to provide the upper bound of contributors who are allowed to observe or access the value \cite{myers1999jflow}. For example, the conclusion of Rule \textsc{T-COND} for conditional expressions may allow contributors from $o_1$,  $o_2$ or $o_3$ to observe the result. However, either the contributors in $o_1 \cup o_2$ or $o_1 \cup o_3$ are allowed at runtime. Such approximation still provides certain security guarantees such that it excludes other sets of observers from observing the value. Moreover, the refined type systems of backends comply with the semantics of the protocols, which provides stronger security guarantees.
\paragraph{Type Refinement and Subtyping}
Harpoon uses labeled types to define values with different accessibility, where $\top$ and $\bot$ denote means sensitive value and public value respectively. In the HIR, types labeled with $\top$ in Harpoon are refined to composite types $\beta$, and types labeled with $\bot$ in Harpoon are refined to atomic types $\alpha$. And the subtype rules follow the ones of Harpoon.

\figref{fig:type-expr} shows the typing rules of expressions. The typing judgment $\TYPEJUDG{\Expr}{t}$ reads expression $e$ has type $t$ under type environment $\Gamma$. A type environment maps variables to types. 
Rule \textsc{T-ARR-RD} allows the array index to be a plain integer or a ciphered integer. If the index is a ciphered integer, then the index has the same set of contributors as the array. 
A similar check applied to the index of Rule \textsc{T-ARR-UPD}. And the array can only be updated with the value who has the same set of contributors as the array.  
Rule \textsc{T-ARR-SLC} extracts a subset of elements from an array from index $n_1$ until $n_2$, where $n_1$ and $n_2$ are plain text integers.
T-SHARE allows a composite value with type $(\alpha, o')$ to be shared among a set of players $o_2$ with threshold $n$ and to be observed by $o_1$. Note that $o'$ is the set of contributors who provide the value, and is dropped in the result type as this information is not needed anymore. 
Rule \textsc{T-RESHARE} allows a shared value to be re-shared to another set of players whose size is greater than the threshold. 
Rule \textsc{T-EVAL} allows the value to be observed by $o'$ if they are a subset of its contributors $o$. 
Rule \textsc{T-UOP} is trivial. Rule \textsc{T-BOP} combines contributors from different operands.
Rule \textsc{T-CONCAT} allows shared arrays to be concatenated if they are shared by the same set of players with the same threshold.
Rule \textsc{T-COND} combines contributors of the condition and the two branches. The contributors of the result type overapproximate the result contributors at runtime.
\begin{figure}[t]
\centering
\footnotesize
\begin{mathpar}
\inferrule*[right=t-nat]{
}
{\TYPEJUDG {n} {\keyw{int}} } 

\inferrule*[right=t-true]{
}
{\TYPEJUDG {true} {\keyw{bool}} }

\inferrule*[right=t-false]{
}
{\TYPEJUDG {false} {\keyw{bool}} }

\inferrule*[right=t-var]{\Gamma(x) = t
}
{\TYPEJUDG{x}{t}} 

\inferrule*[right=t-arr-lit]{\TYPEJUDG{\Expr_i}{\alpha} \\ 0< i \leq n}
{\TYPEJUDG{\{\overline{\Expr}\}_n}{[\alpha]_n}}

\inferrule*[right=t-arr-rd]{\TYPEJUDG{x}{([\alpha]_n,o)} \\\\ \TYPEJUDG{\Expr}{\keyw{int} \mbox{ or } (\keyw{int},o)}}
{\TYPEJUDG{x[\Expr]}{(\alpha, o)}}

\inferrule*[right=t-arr-upd]{\TYPEJUDG{x}{(\alpha[n], o)} \\\\ \TYPEJUDG{\Expr_1}{\keyw{int} \mbox{ or } (\keyw{int}, o)} \\\\ \TYPEJUDG{\Expr_2}{(\alpha, o)}}
{\TYPEJUDG {x.\keyw{upd}(\Expr_1, \Expr_2);} {(\alpha[n], o)} }

\inferrule*[right=t-arr-slc]{\TYPEJUDG{x}{([\alpha]_n, o)} \\ n \geq n_2 > n_1 \geq 0}
{\TYPEJUDG {x.\keyw{slc}(n_1, n_2);} {([\alpha]_{n_2 - n_1}, o)}} 

\inferrule*[right=t-share]{\TYPEJUDG{\Expr}{(\alpha, o')}  \\ |o_2| > 1 \\ n > 1}
{\TYPEJUDG{\keyw{share}(\Expr, o_1, o_2, n)}{\keyw{sh}_{(o_2, n)}(\alpha, o_1)}} 

\inferrule*[right=t-reshare]{\TYPEJUDG{\Expr}{\keyw{sh}_{(o_1, n)}(\alpha, o_2)} \\ |o| \geq n}
{\TYPEJUDG{\keyw{reshare}(\Expr, o)}{\keyw{sh}_{(o,n)}(\alpha, o_2)}} 

\inferrule*[right=t-combine]{\TYPEJUDG{\Expr}{\keyw{sh}_{(o',n)}(\alpha, o'')} \\\\ o_1 \subseteq o' \\ |o_1| >= n \\ o_2 \subseteq o''}
{\TYPEJUDG{\keyw{combine}(\Expr, o_1, o_2)}{\alpha}} 

\inferrule*[right=t-eval]{\TYPEJUDG{\Expr}{(\alpha, o)} \\ o' \subseteq o }
{\TYPEJUDG{\keyw{eval}(\Expr, o')}{\alpha}} 

\inferrule*[right=t-uop]{\TYPEJUDG{\Expr}{(\alpha, o)} \\\\ \TYPEJUDG{\Diamond_u}{\alpha \rightarrow \alpha'}}
{\TYPEJUDG{\Diamond_u \Expr}{(\alpha', o)}}

\inferrule*[right=t-bop]{\TYPEJUDG{\Expr_1}{(\alpha_1, o_1)} \\\\ \TYPEJUDG{\Expr_2}{(\alpha_2, o_2)} \\\\ \TYPEJUDG{\Diamond_b}{\alpha_1 \rightarrow \alpha_2 \rightarrow \alpha}}
{\TYPEJUDG{\Expr_1 \Diamond_b \Expr_2}{(\alpha, o_1 \cup o_2)}}

\inferrule*[right=t-bop-sh]{\TYPEJUDG{\Expr_1}{\keyw{sh}_{(o, n)}(\alpha_1, o_1)} \\\\ \TYPEJUDG{\Expr_2}{\keyw{sh}_{(o, n)}(\alpha_2, o_2)} \\\\ \TYPEJUDG{\Diamond_b}{\alpha_1 \rightarrow \alpha_2 \rightarrow \alpha}}
{\TYPEJUDG{\Expr_1 \Diamond_b \Expr_2}{\keyw{sh}_{(o, n)}(\alpha, o_1 \cup o_2)}}

\inferrule*[right=t-concat]{\TYPEJUDG{\Expr_1}{\keyw{sh}_{(o, n)}([\alpha]_{n_1}, o_1)} \\\\ \TYPEJUDG{\Expr_2}{\keyw{sh}_{(o, n)}([\alpha]_{n_2},o_2)}}
{\TYPEJUDG{\Expr_1 ++ \Expr_2}{\keyw{sh}_{(o, n)}([\alpha]_{n_1 + n_2}, o_1 \cup o_2)}}

\inferrule*[right=t-cond]{\TYPEJUDG{\Expr}{(\keyw{bool},o)} \\\\ \TYPEJUDG{\Expr_1}{(\alpha, o_1)} \\\\ \TYPEJUDG{\Expr_2}{(\alpha, o_2)} }
{\TYPEJUDG{\Expr?\Expr_1:\Expr_2} {(\alpha, o \cup o_1 \cup o_2)}} 

\inferrule*[right=t-s-cond-sh]{\TYPEJUDG{\Expr}{\keyw{sh}_{(o, n)}(\keyw{bool},o_1)} \\ \TYPEJUDG{\Expr_1}{\keyw{sh}_{(o, n)}(\alpha, o_1)} \\ \TYPEJUDG{\Expr_2}{\keyw{sh}_{(o, n)}(\alpha, o_2)} }
{\TYPEJUDG{\Expr?\Expr_1:\Expr_2} {\keyw{sh}_{(o, n)}(\alpha, o_1 \cup o_2 \cup o_3)}}
\end{mathpar}
\caption{The typing rules of expressions.}
\label{fig:type-expr}
\end{figure}


\figref{fig:type-stmt} shows the typing rules of statements. The typing judgment $\STMTTYPEJUDG {\Stmt} {\Gamma'}$ reads statement $\Stmt$ is type-checked under typing environment $\Gamma$ and produces a new environment $\Gamma'$. The type rules of statements are not surprising.
\begin{figure}[t]
\centering
\footnotesize
\begin{mathpar}
\inferrule*[right=t-Skip]{
}
{\STMTTYPEJUDG {\keyw{skip};} {\Gamma}}

\inferrule*[right=t-Dcl]
{\TYPEJUDG{\Expr}{t'} \\ x \not\in \DOM(\Gamma) \\ t' \SUBTYPE t}
{\STMTTYPEJUDG{\keyw{val} ~ x : T := e;}{\Gamma, x : t} }

\inferrule*[right=t-Assign]
{\TYPEJUDG{\Expr}{t} \\  x \not\in \DOM(\Gamma)}
{\STMTTYPEJUDG{\keyw{val} ~ x := e;}{\Gamma, x : t}} 

\inferrule*[right=t-Seq]{\TYPEJUDG{\Stmt_1}{\Gamma''} \\ \TYPEJUDGLIST{\Gamma''  }{\Stmt_2}{\Gamma' } }
{\STMTTYPEJUDG{\Stmt_1\Stmt_2}{\Gamma'}}

\end{mathpar}
\caption{The typing rules of statements}
\label{fig:type-stmt}
\end{figure}%

\subsection{Refined type rules for backends}
\label{sec:types-backends}
This section describes the refined type rules for the cryptographic backends.

\subsubsection{FHE}
\label{sec:type-fhe}
We use the PALISADE library \cite{palisade} and the TFHE library \cite{TFHE} as our FHE backends. Both are asymmetric protocols that use a pair of \texttt{Public} and \texttt{Secret} (or \texttt{Private}) keys for encryption and decryption. The following rule of variable declaration statements means that the plaintext $\Expr$ is encrypted by a singleton set of contributor $\{n\}$, who holds a pair of \texttt{Public} and \texttt{Secret} (or \texttt{Private}) keys. 
\[
\begin{array}{l}
\inferrule*[right=t-fhe-dcl]{\STMTTYPEJUDG{\Expr}{\alpha}}
{\STMTTYPEJUDG {\keyw{val} ~ x : (\alpha, \{n\}) := \Expr;} {\Gamma, x : (\alpha, \{n\})} } \\ 
where\ x \not\in \DOM(\Gamma)
\end{array}
\]

\figref{fig:type-expr-fhe} shows the typing rules of expressions of the FHE backend. Note that the expressions \keyw{share}, \keyw{reshare} and \keyw{combine} are not applicable for FHE backend. Thus, shared types $\keyw{sh}_{(o_1, n)}(\alpha, o_2)$ are not introduced. Thus, there is no rules for them. Rule \textsc{T-FHE-EVAL} enforces that only the contributor is allowed to observe the value, which is consistent with the FHE backends so that the one holding the \texttt{Secret} keys can reveal (or decrypt) the ciphertext. Rule \textsc{T-FHE-BOP} enforces binary operations on ciphertexts from the same contributor.
\begin{figure}[t]
\centering
\footnotesize
\begin{mathpar}

\inferrule*[right=t-fhe-arr-rd]{\TYPEJUDG{x}{([\alpha]_n,\{n'\})} \\ \TYPEJUDG{\Expr}{(\keyw{int}, \{n'\}) \mbox{ or } \keyw{int}}}
{\TYPEJUDG{x[\Expr]}{(\alpha, \{n'\})}} 

\inferrule*[right=t-fhe-arr-upd]{\TYPEJUDG{x}{(\alpha[n], \{n'\})} \\ \TYPEJUDG{\Expr_1}{(\keyw{int}, \{n'\}) \mbox{ or } \keyw{int}} \\ \TYPEJUDG{\Expr_2}{(\alpha, \{n'\})}}
{\TYPEJUDG {x.\keyw{upd}(\Expr_1, \Expr_2);} {(\alpha[n], \{n'\})} } 

\inferrule*[right=t-fhe-arr-slc]{\TYPEJUDG{x}{([\alpha]_n, \{n'\})}}
{\TYPEJUDG {x.\keyw{slc}(n_1, n_2);} {([\alpha]_{n_2 - n_1}, \{n'\})}} 

\inferrule*[right=t-fhe-eval]{\TYPEJUDG{\Expr}{(\alpha, \{n\})}}
{\TYPEJUDG{\keyw{eval}(\Expr, \{n\})}{\alpha}} 

\inferrule*[right=t-fhe-uop]{\TYPEJUDG{\Expr}{(\alpha, \{n\})} \\ \TYPEJUDG{\Diamond_u}{\alpha \rightarrow \alpha}}
{\TYPEJUDG{\Diamond_u \Expr}{(\alpha, \{n\})}}

\inferrule*[right=t-fhe-bop]{\TYPEJUDG{\Expr_1}{(\alpha, \{n\})} \\ \TYPEJUDG{\Expr_2}{(\alpha, \{n\})} \\ \TYPEJUDG{\Diamond_b}{\alpha \rightarrow \alpha \rightarrow \alpha}}
{\TYPEJUDG{\Expr_1 \Diamond_b \Expr_2}{(\alpha, \{n\})}}

\inferrule*[right=t-fhe-cond]{\TYPEJUDG{\Expr}{(\keyw{bool},\{n\})} \\ \TYPEJUDG{\Expr_1}{(\alpha, \{n\})} \\ \TYPEJUDG{\Expr_2}{(\alpha, \{n\})} }
{\TYPEJUDG{\Expr?\Expr_1:\Expr_2} {(\alpha, \{n\})}}

\end{mathpar}
\caption{The typing rules of expressions of the FHE backend.}
\label{fig:type-expr-fhe}
\end{figure}%

\subsubsection{Garbled circuits}
\figref{fig:type-expr-gc} shows the typing rules of expressions of garbled circuits.
\begin{figure}[H]
\centering
\footnotesize
\begin{mathpar}
\inferrule*[right=t-fhe-arr-rd]{\TYPEJUDG{x}{([\alpha]_n,\{n'\})} \\ \TYPEJUDG{\Expr}{(\keyw{int}, \{n'\}) \mbox{ or } \keyw{int}}}
{\TYPEJUDG{x[\Expr]}{(\alpha, \{n'\})}} 

\inferrule*[right=t-fhe-arr-upd]{\TYPEJUDG{x}{(\alpha[n], \{n'\})} \\ \TYPEJUDG{\Expr_1}{(\keyw{int}, \{n'\}) \mbox{ or } \keyw{int}} \\\\ \TYPEJUDG{\Expr_2}{(\alpha, \{n'\})}}
{\TYPEJUDG {x.\keyw{upd}(\Expr_1, \Expr_2);} {(\alpha[n], \{n'\})} } 

\inferrule*[right=t-fhe-arr-slc]{\TYPEJUDG{x}{([\alpha]_n, \{n'\})}}
{\TYPEJUDG {x.\keyw{slc}(n_1, n_2);} {([\alpha]_{n_2 - n_1}, \{n'\})}} 

\inferrule*[right=t-fhe-eval]{\TYPEJUDG{\Expr}{(\alpha, \{n\})}}
{\TYPEJUDG{\keyw{eval}(\Expr, \{n\})}{\alpha}} 

\inferrule*[right=t-fhe-uop]{\TYPEJUDG{\Expr}{(\alpha, \{n\})} \\\\ \TYPEJUDG{\Diamond_u}{\alpha \rightarrow \alpha}}
{\TYPEJUDG{\Diamond_u \Expr}{(\alpha, \{n\})}}

\inferrule*[right=t-fhe-bop]{\TYPEJUDG{\Expr_1}{(\alpha, \{n\})} \\ \TYPEJUDG{\Expr_2}{(\alpha, \{n\})} \\ \TYPEJUDG{\Diamond_b}{\alpha \rightarrow \alpha \rightarrow \alpha}}
{\TYPEJUDG{\Expr_1 \Diamond_b \Expr_2}{(\alpha, \{n\})}}

\inferrule*[right=t-fhe-cond]{\TYPEJUDG{\Expr}{(\keyw{bool},\{n\})} \\ \TYPEJUDG{\Expr_1}{(\alpha, \{n\})} \\ \TYPEJUDG{\Expr_2}{(\alpha, \{n\})} }
{\TYPEJUDG{\Expr?\Expr_1:\Expr_2} {(\alpha, \{n\})}}

\end{mathpar}
\caption{The typing rules of expressions of garbled circuits.}
\label{fig:type-expr-gc}
\end{figure}%

\subsubsection{BGW}
We use the VIFF library \cite{vifflib} as our secret sharing backend. It implements the BGW protocol and offers methods for addition, multiplication, etc. These methods operate on shared values.

\figref{fig:type-expr-shamir-ss} shows the typing rules of expressions for the Shamir secret sharing scheme. Rule \textsc{T-S-SHARE} introduces shared types, and checks that the threshold is greater than 1 and the number of players is greater than or equal to the threshold. Rule \textsc{T-S-COMBINE} eliminates shared types, and checks that at least $t$ players together can open the secret and only legitimate players are allowed to see the value. Rule \textsc{T-S-RESHARE} checks that the shared value may be re-shared to a subset of the original players, where the size of the players is greater or equal to the threshold.  Rule \textsc{T-S-BOP}, \textsc{T-SCONCAT} and \textsc{T-S-COND} are similar to those of the HIR in \figref{fig:type-expr} except that the observers of the result type are the intersection of the operands'. They do not invalidate the soundness of the HIR type system, as the HIR type rules approximate the actual observers at runtime.
\begin{figure}[t]
\centering
\footnotesize
\begin{mathpar}
\inferrule*[right=t-s-share]{\TYPEJUDG{\Expr}{(\alpha, o')} \\ |o_2| \geq n \\ n > 1}
{\TYPEJUDG{\keyw{share}(\Expr, o_1, o_2, n)}{\keyw{sh}_{(o_2, n)}(\alpha, o)}} \\

\inferrule*[right=t-s-reshare]{\TYPEJUDG{\Expr}{\keyw{sh}_{(o_1, n)}(\alpha, o_2)} \\ |o| \geq n \\ o \subseteq o_1}
{\TYPEJUDG{\keyw{reshare}(\Expr, o)}{\keyw{sh}_{(o, n)}(\alpha, o_2)}} \\

\inferrule*[right=t-s-combine]{\TYPEJUDG{\Expr}{\keyw{sh}_{(o',n)}(\alpha, o'')} \\ o_1 \subseteq o' \\ |o_1| \geq n \\ o'' \subseteq o_2}
{\TYPEJUDG{\keyw{combine}(\Expr, o_1, o_2)}{\alpha}}

\inferrule*[right=t-s-bop]{\TYPEJUDG{\Expr_1}{\keyw{sh}_{(o, n)}(\alpha_1, o_1)} \\ \TYPEJUDG{\Expr_2}{\keyw{sh}_{(o, n)}(\alpha_2, o_2)} \\ \TYPEJUDG{\Diamond_b}{\alpha_1 \rightarrow \alpha_2 \rightarrow \alpha}}
{\TYPEJUDG{\Expr_1 \Diamond_b \Expr_2}{\keyw{sh}_{(o, n)}(\alpha, o_1 \cap o_2)}}

\inferrule*[right=t-s-concat]{\TYPEJUDG{\Expr_1}{\keyw{sh}_{(o, n)}([\alpha]_{n_1}, o_1)} \\ \TYPEJUDG{\Expr_2}{\keyw{sh}_{(o, n)}([\alpha]_{n_2},o_2)}}
{\TYPEJUDG{\Expr_1 ++ \Expr_2}{\keyw{sh}_{(o, n)}([\alpha]_{n_1 + n_2}, o_1 \cap o_2)}}

\inferrule*[right=t-s-cond]{\TYPEJUDG{\Expr}{\keyw{sh}_{(o, n)}(\keyw{bool},o_1)} \\ \TYPEJUDG{\Expr_1}{\keyw{sh}_{(o, n)}(\alpha, o_1)} \\ \TYPEJUDG{\Expr_2}{\keyw{sh}_{(o, n)}(\alpha, o_2)} }
{\TYPEJUDG{\Expr?\Expr_1:\Expr_2} {\keyw{sh}_{(o, n)}(\alpha, o_1 \cap o_2 \cap o_3)}}

\end{mathpar}
\caption{The typing rules of expressions of the Shamir secret sharing scheme.}
\label{fig:type-expr-shamir-ss}
\end{figure}

\subsubsection{SPDZ}
\figref{fig:type-expr-add-ss} shows the typing rules of expressions for additive secret sharing implemented by SPDZ. Rule \textsc{T-A-SHARE} introduces shared types, and checks that the threshold is the size of the players, as additive secret sharing is a n-out-of-n secret sharing scheme. Rule \textsc{T-A-COMBINE} eliminates shared types, and checks that all the players together can open the secret and only legitimate players are allowed to see the value. 
Rule \textsc{T-A-BOP}, \textsc{T-A-CONCAT} and \textsc{T-A-COND} are similar to those of the HIR in \figref{fig:type-expr} except that the observers of the result type are the intersection of the operands'. They do not invalidate the soundness of the HIR type system, as the HIR type rules approximate the actual observers at runtime.
\begin{figure}[t]
\centering
\footnotesize
\begin{mathpar}
\inferrule*[right=t-a-share]{\TYPEJUDG{\Expr}{(\alpha, o')} \\ |o_2| > 1}
{\TYPEJUDG{\keyw{share}(\Expr, o_1, o_2, |o_2|)}{\keyw{sh}_{(o_2, |o_2|)}(\alpha, o_1)}} 

\inferrule*[right=t-a-reshare]{\TYPEJUDG{\Expr}{\keyw{sh}_{(o', |o'|)}(\alpha, o_1)} \\ |o| > 1}
{\TYPEJUDG{\keyw{reshare}(\Expr, o)}{\keyw{sh}_{(o, |o|)}(\alpha, o_1)}} \\

\inferrule*[right=t-a-combine]{\TYPEJUDG{\Expr}{\keyw{sh}_{(o, |o|)}(\alpha, o')} \\ o_1 \subseteq o'}
{\TYPEJUDG{\keyw{combine}(\Expr, o, o_1)}{\alpha}} 

\inferrule*[right=t-a-bop]{\TYPEJUDG{\Expr_1}{\keyw{sh}_{(o, |o|)}(\alpha_1, o_1)} \\ \TYPEJUDG{\Expr_2}{\keyw{sh}_{(o, |o|)}(\alpha_2, o_2)} \\ \TYPEJUDG{\Diamond_b}{\alpha_1 \rightarrow \alpha_2 \rightarrow \alpha}}
{\TYPEJUDG{\Expr_1 \Diamond_b \Expr_2}{\keyw{sh}_{(o, |o|)}(\alpha, o_1 \cap o_2)}}

\inferrule*[right=t-a-concat]{\TYPEJUDG{\Expr_1}{\keyw{sh}_{(o, |o|)}([\alpha]_{n_1}, o_1)} \\ \TYPEJUDG{\Expr_2}{\keyw{sh}_{(o, |o|)}([\alpha]_{n_2},o_2)}}
{\TYPEJUDG{\Expr_1 ++ \Expr_2}{\keyw{sh}_{(o, |o|)}([\alpha]_{n_1 + n_2}, o_1 \cap o_2)}}

\inferrule*[right=t-a-cond]{\TYPEJUDG{\Expr}{\keyw{sh}_{(o, |o|)}(\keyw{bool},o_1)} \\ \TYPEJUDG{\Expr_1}{\keyw{sh}_{(o, |o|)}(\alpha, o_2)} \\ \TYPEJUDG{\Expr_2}{\keyw{sh}_{(o, |o|)}(\alpha, o_3)} } 
{\TYPEJUDG{\Expr?\Expr_1:\Expr_2} {\keyw{sh}_{(o, |o|)}(\alpha, o_1 \cap o_2 \cap o_3)}}

\end{mathpar}
\caption{The typing rules of expressions of additive secret sharing.}
\label{fig:type-expr-add-ss}
\end{figure}

\begin{figure}[H]
\begin{minipage}[t]{0.5\textwidth}
\begin{lstlisting}[language=Python,basicstyle=\scriptsize\ttfamily]
def auction():
  x0 = sint(15) # owned by 2 
  x1 = sint(12) # owned by 0 
  x2 = sint(20) # owned by 1 
  x3 = x1<x2 # owned by 0 1 
  x4 = x3 * x1 + (1 - x3) * x2 # owned by 0 1 
  x5 = x0<x4 # owned by 0 1 2 
  x6 = x5 * x4 + (1 - x5) * x0 # owned by 0 1 2 
  x6.reveal_to(0)
  x6.reveal_to(1)
  x6.reveal_to(2) # owned by 
  x10 = sint(2) # owned by 2 
  x7 = sint(1) # owned by 1 
  x8 = sint(0) # owned by 0 
  x9 = x3 * x7 + (1 - x3) * x8 # owned by 0 1 
  x11 = x5 * x9 + (1 - x5) * x10 # owned by 0 1 2 
  x11.reveal_to(0)
  x11.reveal_to(1)
  x11.reveal_to(2) 
\end{lstlisting}
\caption{Generate SPDZ program that computes a second-price auction. For simplicity, the program uses plaintext values as input instead of obtaining them at runtime. There are three bids (15, 12 and 20) whose holders are denoted as 0, 1 and 2 respectively.}
\label{fig:spdz-auction}
\end{minipage}
\begin{minipage}[t]{0.5\textwidth}
\begin{lstlisting}[language=C,basicstyle=\scriptsize\ttfamily]
void TFHE_Addition(TFheGateBootstrappingCloudKeySet* x0, 
  TFheGateBootstrappingParameterSet* x1){
  TFheGateBootstrappingSecretKeySet* x2 = 
    new_random_gate_bootstrapping_secret_keyset(x1);
  x0 = &x2->cloud;
  LweSample* x3 = 
    new_gate_bootstrapping_ciphertext_array(64, x2->params);
  for(uint64_t i = 0; i < 64; i ++) {
    bootsSymEncrypt(&x3[i], ((uint64_t) value>>i)&10, x2);
  }
  LweSample* x4 = 
    new_gate_bootstrapping_ciphertext_array(64, x2->params);
  for(uint64_t i = 0; i < 64; i ++) {
    bootsSymEncrypt(&x4[i], ((uint64_t) value>>i)&5, x2);
  }
  LweSample* x5 = 
    new_gate_bootstrapping_ciphertext_array(64, x2->params);
  fhe_add(x5, a, b, 64, bk);
  printf("%ld\n", num_eval(x5, 64, x2));
}
\end{lstlisting}
\caption{Generated the TFHE program that perforom a secure addition of two 64-bit integers (10 and 5). For simplicity, the program uses plaintext values as input instead of obtaining them at runtime.}
\label{fig:add-tfhe}
\end{minipage}
\begin{minipage}[t]{0.5\textwidth}
\begin{lstlisting}[language=C,basicstyle=\scriptsize\ttfamily]
/************* Functions **************/
static inline Plaintext num_eval(CryptoContext<Poly> cc, 
  const LPPrivateKey<Poly> pk, 
  const Ciphertext<Poly> value) {
  Plaintext ret;
  cc->Decrypt(pk, value, &ret);
  return ret;
}
/**************** BGV_Addition ****************/
void BGV_Addition(CryptoContext<Poly> x0){
  LPKeyPair<Poly> x2 = x0->KeyGen();
  x0->EvalMultKeyGen(x2.secretKey);
  cout << num_eval(x0, x2.secretKey, 
    x0->EvalAdd(x0->Encrypt(x2.publicKey, 
    x0->MakeFractionalPlaintext(10)),
    x0->Encrypt(x2.publicKey, x0->MakeFractionalPlaintext(5))
  )) << endl;
 }
\end{lstlisting}
\caption{Generated the HElib program that perforom a secure addition of two integers (10 and 5). For simplicity, the program uses plaintext values as input instead of obtaining them at runtime.}
\label{fig:add-bgv}
\end{minipage}
\end{figure}

\begin{figure}[H]
\begin{lstlisting}[language=Python,basicstyle=\scriptsize\ttfamily,morekeywords={async, await}]
from honeybadgermpc.progs.mixins.share_comparison 
  import Equality, LessThan
from honeybadgermpc.preprocessing import 
  (PreProcessedElements as FakePreProcessedElements,)
from honeybadgermpc.progs.mixins.share_arithmetic import 
  (BeaverMultiply, BeaverMultiplyArrays, MixinConstants, )
import logging
import sys
import asyncio
from time import time
mpc_config = { 
  MixinConstants.MultiplyShareArray: BeaverMultiplyArrays(), 
  MixinConstants.MultiplyShare: BeaverMultiply(), 
  MixinConstants.ShareLessThan: LessThan(),}

async def Auction(x0): # x0: an MPC context
  x2 = x0.ShareFuture()
  x2.ret.set_result(Share(12))
  x3 = x0.ShareFuture()
  x3.ret.set_result(Share(20))
  x4 = x0.ShareFuture()
  x4.ret.set_result(Share(15))
  x5 = (x2 < x3)
  x6 = x5 * x2 + (1 - x5) * x3
  x7 = (x4 < x6)
  print(await (x7 * x6 + (1 - x7) * x4).open())
  print(await (x7 * (x5 * num(x0,1) + (1 - x5) * num(x0,0)) 
    + (1 - x7) * num(x0,2)).open())

async def _run(peers, n, t, my_id, k): 
  from honeybadgermpc.ipc import ProcessProgramRunner
  async with ProcessProgramRunner(
    peers, n, t, my_id, mpc_config) as runner:
    await runner.execute("0", Auction)
    bytes_sent = runner.node_communicator.bytes_sent
    print(f"[my_id] Total bytes sent out: bytes_sent")
  
if __name__ == "__main__": 
  from honeybadgermpc.config import HbmpcConfig
  import sys
  HbmpcConfig.load_config()
  
  asyncio.set_event_loop(asyncio.new_event_loop())
  loop = asyncio.get_event_loop()
  loop.set_debug(False)
  try: 
    pp_elements = FakePreProcessedElements()
    k = 3  
    if HbmpcConfig.my_id == 0: 
      num_bits = 40
      pp_elements.generate_triples(num_bits * 20 * k, 
        HbmpcConfig.N, HbmpcConfig.t)
      pp_elements.generate_share_bits(num_bits * k, 
        HbmpcConfig.N, HbmpcConfig.t)
      pp_elements.preprocessing_done()
    
    else: 
      loop.run_until_complete(
        pp_elements.wait_for_preprocessing())
    
    loop.run_until_complete(
    _run(HbmpcConfig.peers, HbmpcConfig.N, 
      HbmpcConfig.t, HbmpcConfig.my_id, k)
    )
  
  finally: 
    loop.close()
    
\end{lstlisting}
\caption{Generate HoneyBadgerMPC that computes a second-price auction. For simplicity, the program uses plaintext values as input instead of obtaining them at runtime. There are three bids (15, 12 and 20) whose holders are denoted as 0, 1 and 2 respectively. For testing and development, the HoneyBadgerMPC program runs in a single-process simulated network.}
\label{fig:hbmpc-auction}
\end{figure}

\begin{figure}[t]
\begin{lstlisting}[language=C,basicstyle=\scriptsize\ttfamily]
#include <helib/FHE.h>
#include <iostream>
#include <stdlib.h>
#include <stdio.h>
#include <stdint.h>
#include <stdbool.h>
using namespace std;
/************* Functions **************/
static inline NTL::ZZX unum_eval(Ctxt n, 
  FHESecKey secretKey) {
  NTL::ZZX ret;
  secretKey.Decrypt(ret, n);
  return ret;
}
static inline Ctxt unum_add(Ctxt n1, Ctxt n2) {
  Ctxt ret = n1;
  ret += n2;
  return ret;
}
static inline Ctxt unum_init(int value, 
  const FHEPubKey& publicKey) {
  Ctxt ret(publicKey);
  publicKey.Encrypt(ret, NTL::to_ZZX(value));
  return ret;
}
static inline Ctxt unum_mul_plain(Ctxt n1, int n2) {
  Ctxt ret = n1;
  ret.multByConstant(NTL::to_ZZX(n2));
  return ret;
}
/**************** Multiply ****************/
void multiply(FHEcontext& x0, int x1){
  FHESecKey x2(x0);
  x2.GenSecKey();
  cout  << unum_eval(unum_add(unum_add(
    unum_init(1, x2), unum_mul_plain(unum_init(2, x2), 2)), 
    unum_init(3, x2)), x2) << 
    unum_eval(unum_add(unum_add(unum_init(4, x2), 
    unum_mul_plain(unum_init(5, x2), 2)), 
    unum_init(6, x2)), x2) << endl;
}
/*****************************************
End of C Generated Code
*******************************************/
int main(){
  long m = 0;                   // Specific modulus
  long p = 1021;                // Plaintext base [default=2], should be a prime number
  long r = 1;                   // Lifting [default=1]
  long L = 16;                  // Number of levels in the modulus chain [default=heuristic]
  long c = 3;                   // Number of columns in key-switching matrix [default=2]
  long w = 64;                  // Hamming weight of secret key
  long d = 0;                   // Degree of the field extension [default=1]
  long k = 128;                 // Security parameter [default=80]
  long s = 0;                   // Minimum number of slots [default=0]
  m = FindM(k, L, c, p, d, s, 0);
  FHEcontext context(m, p, r);
  buildModChain(context, L, c);
  NTL::ZZX  G =  context.alMod.getFactorsOverZZ()[0];
  multiply(context, 0);
  return 0;
}

\end{lstlisting}
\caption{Generated HElib program that computes matrix-vector multiplication of \texttt{Array(Array(1, 2, 3), Array(4, 5, 6))} and  \texttt{Array(1, 2, 1)}. For simplicity, the program uses plaintext values as input instead of obtaining them at runtime.}
\label{fig:helib-mul}
\end{figure}

\begin{figure}[t]
\begin{lstlisting}[language=C,basicstyle=\scriptsize\ttfamily]
#include <stdlib.h>
#include <tfhe/tfhe.h>
#include "ir.h"
#include <stdio.h>
#include <stdint.h>
#include <stdbool.h>
static inline LweSample* add(const LweSample* a, 
  const LweSample* b, uint64_t size, 
  const TFheGateBootstrappingCloudKeySet* bk) {
  LweSample* ret = 
    new_gate_bootstrapping_ciphertext_array(size, bk->params);
  fhe_add(ret, a, b, size, bk);
  return ret;
}
static inline LweSample* unum_init(uint64_t value, 
  uint64_t size, 
  const TFheGateBootstrappingSecretKeySet* bk) {
  LweSample* ret = 
    new_gate_bootstrapping_ciphertext_array(size, bk->params);
  for(uint64_t i = 0; i < size; i++) {
    bootsSymEncrypt(&ret[i], ((uint64_t) value>>i)&1, bk);
  }
  return ret;
}
static inline LweSample* unum_mul_plain(const LweSample* a, 
  const int b, uint64_t size, 
  const TFheGateBootstrappingCloudKeySet* bk) {
  LweSample* c_b = 
    new_gate_bootstrapping_ciphertext_array(size, bk->params);
  for(uint64_t i = 0; i < size; i ++) {
    bootsCONSTANT(&c_b[i], ((uint64_t) b>>i)&1, bk);
  }
  LweSample* ret = 
    new_gate_bootstrapping_ciphertext_array(size, bk->params);
  fhe_mul(ret, a, c_b, size, bk);
  return ret;
}
static inline uint64_t unum_eval(const LweSample* a, 
  uint64_t size, 
  const TFheGateBootstrappingSecretKeySet* key) {
  assert(size <= 64);
  uint64_t res[size];
  uint64_t ret = 0;
  for(uint64_t i=0; i<size; i++){
    res[i] = bootsSymDecrypt(&a[i], key)>0;
  }
  for(uint64_t i=0; i<size; i++){
    ret |= ((uint64_t) res[i])<<((uint64_t) i);
  }
  return ret;
}
void multiply(TFheGateBootstrappingCloudKeySet* x0, 
  TFheGateBootstrappingParameterSet* x1){
  TFheGateBootstrappingSecretKeySet* x2 = 
    new_random_gate_bootstrapping_secret_keyset(x1);
  x0 = &x2->cloud;
  const LweSample** x3 = 
    (const LweSample**)malloc(2 * sizeof(const LweSample*));
  LweSample* x4 = add(unum_init(1, 64 ,x2), x3[0], 64 , x0);
  x3[0] = x4;
  LweSample* x5 = add(unum_mul_plain(
    unum_init(2, 64 ,x2), 2, 64, x0), x4, 64 , x0);
  x3[0] = x5;
  LweSample* x6 = add(unum_init(3, 64 ,x2), x5, 64 , x0);
  x3[0] = x6;
  printf("%ld\n", unum_eval(x6, 64, x2));
  x3[1] = add(unum_init(6, 64 ,x2), 
    add(unum_mul_plain(unum_init(5, 64 ,x2), 2, 64, x0), 
    add(unum_init(4, 64 ,x2), x3[1], 64, x0), 64, x0), 64, x0);
}
\end{lstlisting}
\caption{Generated TFHE code that computes matrix-vector multiplication of \texttt{Array(Array(1, 2, 3), Array(4, 5, 6))} and  \texttt{Array(1, 2, 1)}. For simplicity, the program uses plaintext values as input instead of obtaining them at runtime. (part1)}
\label{fig:tfhe-matrxi-vec1}
\end{figure}

\begin{figure}[t]
\begin{minipage}[t]{0.5\textwidth}
\begin{lstlisting}[language=C,basicstyle=\scriptsize\ttfamily]
int main(){
  const int minimum_lambda = 110;
  TFheGateBootstrappingParameterSet* params = 
    new_default_gate_bootstrapping_parameters(minimum_lambda);
  uint32_t seed[] = { 314, 1592, 657 };
  tfhe_random_generator_setSeed(seed,3);
  TFheGateBootstrappingSecretKeySet* s_key = 
    new_random_gate_bootstrapping_secret_keyset(params);
  const TFheGateBootstrappingCloudKeySet* c_key = &s_key->cloud;
  multiply(NULL, params);
  return 0;
}
\end{lstlisting}
\caption{Generated TFHE code that computes matrix-vector multiplication of \texttt{Array(Array(1, 2, 3), Array(4, 5, 6))} and  \texttt{Array(1, 2, 1)}. For simplicity, the program uses plaintext values as input instead of obtaining them at runtime. (part2)}
\label{fig:tfhe-matrxi-vec2}
\end{minipage}
\begin{minipage}[t]{0.5\textwidth}
\begin{lstlisting}[language=C,basicstyle=\scriptsize\ttfamily]
#include <assert.h>
#include <stdlib.h>
#include <tfhe/tfhe.h>
#include "ir.h"
#include <stdio.h>
#include <stdint.h>
#include <stdbool.h>
#include <stdarg.h>
/************* Functions **************/
static inline LweSample* add(const LweSample* a, 
  const LweSample* b, uint64_t size, 
  const TFheGateBootstrappingCloudKeySet* bk) {
  LweSample* ret = 
    new_gate_bootstrapping_ciphertext_array(size, bk->params);
  fhe_add(ret, a, b, size, bk);
  return ret;
}
LweSample* num_array_init(const uint64_t bit_width, 
  const TFheGateBootstrappingSecretKeySet* bk, 
  const uint64_t size, ...){
  va_list valist;
  va_start(valist, size);
  LweSample* ret = 
    new_gate_bootstrapping_ciphertext_array(bit_width * size, 
      bk->params);
  for (uint64_t i = 0; i < size ; i ++) {
    uint64_t tmp = va_arg(valist, int);
    for(uint64_t j = 0; j < bit_width; j ++) {
      bootsSymEncrypt(&ret[i * bit_width + j], 
        ((uint64_t) tmp>>j)&1, bk);
    }
  }
  va_end(valist);
  return ret;
}
static inline LweSample* num_init(uint64_t value, uint64_t size, 
  const TFheGateBootstrappingSecretKeySet* bk) {
  LweSample* ret = 
    new_gate_bootstrapping_ciphertext_array(size, bk->params);
  for(uint64_t i = 0; i < size; i ++) {
    bootsSymEncrypt(&ret[i], ((uint64_t) value>>i)&1, bk);
  }
  return ret;
}
static inline LweSample* num_leq(const LweSample* a, 
  const LweSample* b, uint64_t size, 
  const TFheGateBootstrappingCloudKeySet* bk) {
  LweSample* ret = 
    new_gate_bootstrapping_ciphertext(bk->params);
  fhe_compare_signed(ret, a, b, size, bk);
  return ret;
}
\end{lstlisting}
\caption{Generated TFHE program that merge sort an array of [3, 1, 5, 2]. For simplicity, the program uses plaintext values as input instead of obtaining them at runtime. (part 1)}
\label{fig:tfhe-merge1}
\end{minipage}
\end{figure}

\begin{figure}[H]
\begin{lstlisting}[language=C,basicstyle=\scriptsize\ttfamily]
static inline LweSample* bit_and(const LweSample* a, 
  const LweSample* b, 
  const TFheGateBootstrappingCloudKeySet* bk) {
  LweSample* ret = new_gate_bootstrapping_ciphertext(bk->params);
  bootsAND(ret, a, b, bk);
  return ret;
}
static inline LweSample* num_array_index(const LweSample* i, 
  const LweSample* arr, uint64_t length, uint64_t bit_width, 
    const TFheGateBootstrappingCloudKeySet* bk) {
  LweSample * res = 
    new_gate_bootstrapping_ciphertext_array(bit_width, bk->params);
  fhe_array_index(res, i, arr, length, bit_width, bk);
  return res;
}
static inline LweSample* num_less(const LweSample* a, 
  const LweSample* b, uint64_t size, 
  const TFheGateBootstrappingCloudKeySet* bk) {
  LweSample* ret = 
    new_gate_bootstrapping_ciphertext(bk->params);
  LweSample* t1 = 
    new_gate_bootstrapping_ciphertext_array(size, bk->params);
  LweSample* t2 = 
    new_gate_bootstrapping_ciphertext_array(size, bk->params);
  fhe_neg(t1, b, size, bk);
  fhe_add(t2, a, t1, size, bk);
  bootsCOPY(ret, &t2[size-1], bk);
  return ret;
}
static inline LweSample* num_mux(const LweSample* a, 
  const LweSample* b, const LweSample* c, uint64_t size, 
  const TFheGateBootstrappingCloudKeySet* bk) {
  LweSample* ret = 
    new_gate_bootstrapping_ciphertext_array(size, 
    bk->params);
  fhe_mux(ret, a, b, c, size, bk);
  return ret;
}
static inline int64_t num_eval(const LweSample* a, 
  uint64_t size, 
  const TFheGateBootstrappingSecretKeySet* key) {
  assert(size <= 64);
  uint64_t res[size];
  int64_t ret = 0;
  for(uint64_t i=0; i<size; i++){
    res[i] = bootsSymDecrypt(&a[i], key)>0;
  }
  if(res[size-1]!=0){
    for(uint64_t i=0; i<size; i++){
      res[i] = 1 - res[i];
      ret |= ((uint64_t) res[i]<<((uint64_t) i));
    }
    ret = -(ret + 1);
  }
  else{
    for(uint64_t i=0; i<size; i++){
      ret |= ((uint64_t) res[i]<<((uint64_t) i));
    }
  }
  return ret;
}
/**************** MergeSort ****************/
void MergeSort(TFheGateBootstrappingCloudKeySet* x0, 
  TFheGateBootstrappingParameterSet* x1){
  TFheGateBootstrappingSecretKeySet* x2 = 
    new_random_gate_bootstrapping_secret_keyset(x1);
  x0 = &x2->cloud;
  LweSample* x3 = num_array_init(16, x2, 2, 5 ,8);
  LweSample* x4 = num_array_init(16, x2, 2, 2 ,4);
  LweSample* x5 = num_init(0, 16, x2);
  LweSample* x6 = num_init(0, 16, x2);
\end{lstlisting}
\caption{Generated TFHE program that merge sort an array of [3, 1, 5, 2]. For simplicity, the program uses plaintext values as input instead of obtaining them at runtime. (part 2)}
\label{fig:tfhe-merge2}
\end{figure}

\begin{figure}[H]
\begin{lstlisting}[language=C,basicstyle=\scriptsize\ttfamily]
LweSample* x7 = num_leq(
  num_array_index(x5, x3, 2, 16, x0), 
  num_array_index(x6, x4, 2, 16, x0), 16 , x0);
const LweSample* x8 = 
  num_less(x5, num_init(2, 16 ,x2), 16 , x0);
const LweSample* x9 = 
  num_less(x6, num_init(2, 16 ,x2), 16 , x0);
LweSample* x10 = bit_and(x8, x9, x0);
LweSample* x11 = num_mux(num_mux(add(x5, 
  num_init(1, 16 ,x2), 16, x0), x5, x7, 16 , x0), 
  num_mux(add(x5, num_init(1, 16 ,x2), 16, x0), 
    x5, x8, 16 , x0), x10, 16 , x0);
LweSample* x12 = num_mux(num_mux(x6, add(x6, 
  num_init(1, 16 ,x2), 16, x0), x7, 16 , x0), 
  num_mux(add(x6, num_init(1, 16 ,x2), 16, x0), 
    x6, x9, 16 , x0), x10, 16 , x0);
LweSample* x13 = num_leq(
  num_array_index(x11, x3, 2, 16, x0), 
  num_array_index(x12, x4, 2, 16, x0), 16 , x0);
const LweSample* x14 = 
  num_less(x11, num_init(2, 16 ,x2), 16 , x0);
const LweSample* x15 = 
  num_less(x12, num_init(2, 16 ,x2), 16 , x0);
LweSample* x16 = bit_and(x14, x15, x0);
LweSample* x17 = num_mux(num_mux(
  add(x11, num_init(1, 16 ,x2), 16, x0), 
  x11, x13, 16 , x0), num_mux(
  add(x11, num_init(1, 16 ,x2), 
  16, x0), x11, x14, 16 , x0), x16, 16 , x0);
LweSample* x18 = 
  num_mux(num_mux(x12, 
  add(x12, num_init(1, 16 ,x2), 16, x0), 
  x13, 16 , x0), num_mux(add(x12, num_init(1, 16 ,x2), 
  16, x0), x12, x15, 16 , x0), x16, 16 , x0);
LweSample* x19 = 
  num_leq(num_array_index(x17, x3, 2, 16, x0), 
  num_array_index(x18, x4, 2, 16, x0), 16 , x0);
const LweSample* x20 = 
  num_less(x17, num_init(2, 16 ,x2), 16 , x0);
const LweSample* x21 = 
  num_less(x18, num_init(2, 16 ,x2), 16 , x0);
LweSample* x22 = bit_and(x20, x21, x0);
LweSample* x23 = num_mux(num_mux(add(x17, 
  num_init(1, 16 ,x2), 16, x0), x17, x19, 16 , x0), 
  num_mux(add(x17, num_init(1, 16 ,x2), 
  16, x0), x17, x20, 16 , x0), x22, 16 , x0);  
LweSample* x24 = num_mux(num_mux(x18, 
  add(x18, num_init(1, 16 ,x2), 16, x0), x19, 16 , x0), 
  num_mux(add(x18, num_init(1, 16 ,x2), 16, x0), 
  x18, x21, 16 , x0), x22, 16 , x0);
const LweSample* x25 = 
  num_less(x23, num_init(2, 16 ,x2), 16 , x0);
printf("%ld\n", num_eval(num_mux(num_mux(
  num_array_index(x5, x3, 2, 16, x0), 
  num_array_index(x6, x4, 2, 16, x0), x7, 16 , x0), 
  num_mux(num_array_index(x5, x3, 2, 16, x0), 
  num_array_index(x6, x4, 2, 16, x0), x8, 16 , x0), 
  x10, 16 , x0), 16, x2));
printf("%ld\n", num_eval(num_mux(num_mux(
  num_array_index(x11, x3, 2, 16, x0), 
  num_array_index(x12, x4, 2, 16, x0), x13, 16 , x0), 
  num_mux(num_array_index(x11, x3, 2, 16, x0), 
  num_array_index(x12, x4, 2, 16, x0), x14, 16 , x0), 
  x16, 16 , x0), 16, x2));
printf("%ld\n", num_eval(num_mux(num_mux(
  num_array_index(x17, x3, 2, 16, x0), 
  num_array_index(x18, x4, 2, 16, x0), x19, 16 , x0), 
  num_mux(num_array_index(x17, x3, 2, 16, x0), 
  num_array_index(x18, x4, 2, 16, x0), x20, 16 , x0), 
  x22, 16 , x0), 16, x2));  
\end{lstlisting}
\caption{Generated TFHE program that merge sort an array of $[3, 1, 5, 2]$. For simplicity, the program uses plaintext values as input instead of obtaining them at runtime. (part 3)}
\label{fig:tfhe-merge3}
\end{figure}

\begin{figure}[H]
\begin{lstlisting}[language=C,basicstyle=\scriptsize\ttfamily]
printf("%ld\n", num_eval(num_mux(num_mux(
  num_array_index(x23, x3, 2, 16, x0), 
  num_array_index(x24, x4, 2, 16, x0), 
  num_leq(num_array_index(x23, x3, 2, 16, x0), 
  num_array_index(x24, x4, 2, 16, x0), 16 , x0), 16 , x0), 
  num_mux(num_array_index(x23, x3, 2, 16, x0), 
  num_array_index(x24, x4, 2, 16, x0), x25, 16 , x0), 
  bit_and(x25, num_less(x24, num_init(2, 16 ,x2), 16 , x0), 
  x0), 16 , x0), 16, x2));
}
int main(){
  const int minimum_lambda = 110;
  TFheGateBootstrappingParameterSet* params = 
    new_default_gate_bootstrapping_parameters(minimum_lambda);
  uint32_t seed[] = { 314, 1592, 657 };
  tfhe_random_generator_setSeed(seed,3);
  TFheGateBootstrappingSecretKeySet* s_key = 
    new_random_gate_bootstrapping_secret_keyset(params);
  const TFheGateBootstrappingCloudKeySet* c_key = &s_key->cloud;
  MergeSort(NULL, params);
  return 0;
}
\end{lstlisting}
\caption{Generated TFHE program that merge sort an array of [3, 1, 5, 2]. For simplicity, the program uses plaintext values as input instead of obtaining them at runtime. (part 4)}
\label{fig:tfhe-merge4}
\end{figure}
}{}

\end{document}